\documentclass{IEEEtran}

\usepackage{booktabs}
\usepackage{multirow}
\usepackage[table]{xcolor}

\usepackage{amsmath,amsfonts}
\usepackage{array}
\usepackage[caption=false,font=normalsize,labelfont=sf,textfont=sf]{subfig}
\usepackage{textcomp}
\usepackage{stfloats}
\usepackage{verbatim}
\usepackage{graphicx}
\usepackage{cite}
\usepackage{tabularx}
\usepackage{booktabs}

\usepackage{acronym}
\acrodef{MIMO}{multiple-input multiple-output}
\acrodef{RF}{radio frequency}
\acrodef{LoS}{line-of-sight}
\acrodef{NLoS}{non-line-of-sight}
\acrodef{AoA}{angle-of-arrival}
\acrodef{AoD}{angle-of-departure}
\acrodef{UPA}{uniform planar array}
\acrodef{ARV}{array response vector}
\acrodef{EM}{electromagnetic}
\acrodef{ER-FAS}{electromagnetically reconfigurable fluid antenna system}
\acrodef{SV}{Saleh-Valenzuela}
\usepackage{color} 
\newcommand{\red}[1]{{\color{red}{#1}}} 
 
\newcommand{\blue}[1]{{\color{blue}{#1}}}

\usepackage{tikz,pgfplots}
\usepackage{amsmath,amssymb}
\pgfplotsset{compat=newest}
\usepackage{tabu,longtable}
\usepackage{diagbox}
\usepackage{threeparttable}
\usepackage{makecell}
\usepackage{multirow} % enable table multirow
\usepackage[bottom=0.75in,top=0.75in,left=0.625in,right=0.625in]{geometry}
\usepackage{units} 		% use \unit command
\usepackage{bm} 		% for bold symbol
\usepackage{amssymb} 	% for AMS symbols
\newcommand{\TT}{\mathsf{T}}
\newcommand{\HH}{\mathsf{H}}

\newcommand{\av}{{\bf a}}
\newcommand{\bv}{{\bf b}}

\newcommand{\dv}{{\bf d}}

\newcommand{\fv}{{\bf f}}
\newcommand{\gv}{{\bf g}}

\newcommand{\kv}{{\bf k}}

\newcommand{\nv}{{\bf n}}

\newcommand{\pv}{{\bf p}}

\newcommand{\rv}{{\bf r}}

\newcommand{\wv}{{\bf w}}

\newcommand{\Bm}{{\bf B}}

\newcommand{\Dm}{{\bf D}}

\newcommand{\Hm}{{\bf H}}

\newcommand{\It}{{\rm I}}

\newcommand{\Rt}{{\rm R}}

\newcommand{\Tt}{{\rm T}}

\newcommand{\Xt}{{\rm X}}

\newcommand{\phiv}{\hbox{\boldmath$\phi$}}
\newcommand{\varphiv}{\hbox{\boldmath$\varphi$}}

\newcommand{\thetav}{\hbox{$\boldsymbol\theta$}}

\usepackage{hyperref} 	% for URL
\usepackage{amsmath}

\usepackage{algorithm}
\usepackage{algpseudocode}% http://ctan.org/pkg/algorithmicx
\algtext*{EndWhile}% Remove "end while" text
\algtext*{EndIf}% Remove "end if" text
\algtext*{EndFor}

\makeatletter
\algnewcommand{\LineComment}[1]{\Statex \hskip\ALG@thistlm \(\triangleright\) #1}
\makeatother
\begin{document}
\bstctlcite{IEEEexample:BSTcontrol}

\title{Electromagnetically Reconfigurable Fluid Antenna System for Wireless Communications: Design, Modeling, Algorithm, Fabrication, and Experiment}

\author{Ruiqi Wang, 
Pinjun Zheng,~\IEEEmembership{Member,~IEEE},
Vijith Varma Kotte, 
Sakandar Rauf, \\
Yiming Yang,
Muhammad Mahboob Ur Rahman,~\IEEEmembership{Senior~Member,~IEEE}, \\
Tareq Y. Al-Naffouri,~\IEEEmembership{Fellow,~IEEE}, 
Atif Shamim,~\IEEEmembership{Fellow,~IEEE} 
\thanks{R. Wang, V. V. Kotte, Y. Yang, M. M. U. Rahman, T. Y. Al-Naffouri, and A. Shamim are with the  Electrical and Computer Engineering Program, Computer, Electrical and Mathematical Sciences and Engineering (CEMSE), King Abdullah University of Science and Technology (KAUST), Thuwal 23955-6900, Kingdom of Saudi Arabia (e-mail: \{ruiqi.wang.1; vijith.kotte; yiming.yang; muhammad.rahman; tareq.alnaffouri; atif.shamim\}@kaust.edu.sa).

P. Zheng is with the School of Engineering, The University of British Columbia, Kelowna, BC V1V 1V7, Canada (e-mail: pinjun.zheng@ubc.ca). The majority of his contributions to this work were made during his Ph.D. studies at KAUST, Thuwal 23955-6900, Kingdom of Saudi Arabia.

S. Rauf is with the School of Biochemistry and Biotechnology, University of the Punjab, Lahore, Pakistan. (e-mail: sakandar.sbb@pu.edu.pk). The majority of his contributions to this work were made at KAUST, Thuwal 23955-6900, Kingdom of Saudi Arabia.
}
}

%\markboth{This work has been submitted to the IEEE for possible publication.  Copyright may be transferred without notice.}{draft}
\maketitle

\begin{abstract}
 
This paper presents the concept, design, channel modeling, beamforming algorithm development, prototype fabrication, and experimental measurement of an electromagnetically reconfigurable fluid antenna system (ER-FAS), in which each FAS array element features electromagnetic (EM) reconfigurability. Unlike most existing FAS works that investigate spatial reconfigurability by adjusting the position and/or orientation of array elements, the proposed ER-FAS enables direct control over the EM characteristics of each element, allowing for dynamic radiation pattern reconfigurability. Specifically, a novel ER-FAS architecture leveraging software-controlled fluidics is proposed, and corresponding wireless channel models are established. Based on this ER-FAS channel model, a low-complexity greedy beamforming algorithm is developed to jointly optimize the analog phase shift and the radiation state of each array element. The accuracy of the ER-FAS channel model and the effectiveness of the beamforming algorithm are validated through (i) full-wave EM simulations and (ii) numerical spectral efficiency evaluations. These results confirm that the proposed ER-FAS significantly enhances spectral efficiency in both near-field and far-field scenarios compared to conventional antenna arrays. To further validate this design, we fabricate prototypes for both the ER-FAS element and array, using Galinstan liquid metal alloy, fluid silver paste, and software-controlled fluidic channels. The simulation results are experimentally validated through prototype measurements conducted in an anechoic chamber. Additionally, several indoor communication experiments using a pair of software-defined radios demonstrate the superior received power and bit error rate performance of the ER-FAS prototype. This paper offers a comprehensive demonstration of a liquid-based ER-FAS array for wireless communication, incorporating a novel electromagnetically reconfigurable design, channel modeling, and beamforming, supported by simulation, hardware implementation, and experimental validation.

\end{abstract}

\begin{IEEEkeywords}

electromagnetically reconfigurable antennas,
fluid antenna system,
wireless communication,
hardware fabrication,
experimental measurement.

\end{IEEEkeywords}

\section{Introduction}

The fifth-generation (5G) and sixth-generation (6G) networks demand extremely high data rates and capacity, necessitating fundamental physical layer innovations~\cite{Dang2020Should,WKK2023FAS_Part2}. Fluid antenna systems (FAS) have recently emerged as a cutting-edge solution to enhance physical layer flexibility, showing great promise for 6G~\cite{NWK2024FAS_Tutorial}. Extensive studies have explored spatially reconfigurable antenna switching for wireless applications~\cite{WKK2021First_Paper_on_FAS, ChaoWang2024FAS, NWK2025Oversampling}. Unlike traditional antennas, FAS enables activation of optimal positions/ports, offering a new degree of freedom to boost spatial diversity and multiplexing~\cite{WKK2023FAS_Part1, WKK2023FAS_Part3}. Moreover, FAS can be integrated with advanced technologies such as reconfigurable intelligent surfaces~\cite{Wang_ruiqi2024TAP}, massive \ac{MIMO}\cite{Ying2024Reconfigurable}, and integrated sensing and communications (ISAC)\cite{LA2022ISAC_Tutorial} to further advance wireless networks.

Generally, FAS and other next-generation reconfigurable antenna (NGRA) technologies can be grouped into three types: spatially reconfigurable, electromagnetically reconfigurable, and time-modulated. In the spatial domain, the position or orientation of the antenna can be flexibly adjusted, forming the spatially reconfigurable FAS (SR-FAS), also referred to as movable antenna (MA)~\cite{Zhu_lipeng2024Magzine_MA}. In SR-FAS or MA, the spatial position can be adjusted, while the electromagnetic characteristics remain unchanged. In this case, FAS and MA share the same concept, where communication performance is enhanced by introducing spatial degrees of freedom~\cite{zhu2024historicalreviewfluidantenna}. In the electromagnetic domain, FAS can vary its intrinsic radiation properties by reshaping metallic patterns or dielectric substrates without altering the position~\cite{Song_lingnan2019TAP}. By reconstructing the radiator geometry, properties such as frequency, polarization, and radiation pattern can be controlled~\cite{Rodrigo2014TAP}, forming the electromagnetically reconfigurable FAS (ER-FAS). In the time domain, reconfiguration can also be achieved by introducing time as a fourth dimension~\cite{Zhu_quanjiang2014TAP}.

In the current literature, extensive theoretical studies on FAS have been conducted. For example,\cite{Khammassi2023TWC} proposed an analytical approximation for the FAS channel,\cite{Wang_chao2024AI-empowered} explored AI-based FAS optimization, and~\cite{WKK2020FAS_limit} characterized performance limits. However, most of these works focus solely on spatial reconfigurability, i.e., SR-FAS or MA. Despite the potential of FAS, most studies remain theoretical, lacking full-wave simulations and experimental validations. Only a few works, such as~\cite{dong2024movableantennaprototyping}, demonstrate experimental results, where an MA system was fabricated and shown to achieve spatial performance gains in practical settings. Nevertheless, MA/SR-FAS only adjusts antenna position without modifying electromagnetic radiation properties and typically relies on heavy mechanical platforms, leading to increased design complexity and cost~\cite{dong2024movableantennaprototyping}, thereby limiting applicability. In contrast, ER-FAS offers a more efficient solution, enabling real-time adjustment of radiation characteristics by altering the state of the RF radiator. A pixel-based reconfigurable FAS with high switching speed was introduced in~\cite{Zhang_jichen2025OJAP}, using variable interconnections between pixels~\cite{Lotfi2017TAP}. However, unused pixels may introduce parasitic coupling. To address this, conductive liquid can be used to control the presence of radiating material, effectively mitigating internal coupling within each reconfigurable element. Moreover, liquid metal-based FAS allows mechanical flexibility to accommodate conformal applications. Overall, ER-FAS studies for communication systems remain at an early stage and will be comprehensively explored in this work.

In this work, we practically design a novel ER-FAS for wireless communication systems, where the radiation pattern of each array element can be reconfigured through software-controllable fluidics. The main contributions of this work are summarized as follows:

\subsubsection{Proposing a Novel Practical ER-FAS Design}
We propose a novel practical ER-FAS design. By associating fluid radiators with controllable feeding phases, this design provides an antenna array with extensive degrees of freedom, enhancing wireless communication performance in complex channel conditions.

\subsubsection{Modeling and Beamforming for ER-FAS-Based Communications}
To characterize the newly introduced ER-FAS features, a single-user \ac{MIMO} channel model is developed in both near-field and far-field scenarios, capturing its EM-domain flexibility over conventional arrays. Furthermore, a greedy beamforming algorithm is designed to optimize analog phase shifts and antenna state selection with low complexity.

\subsubsection{Full-Wave Validation of the Derived Model and Numerical Evaluation of the Proposed Beamforming Algorithm}
We perform full-wave simulations to validate the proposed ER-FAS channel model and evaluate the spectral efficiency to assess the effectiveness of the beamforming algorithm. The results show the superior spectral efficiency of the ER-FAS design compared to conventional antenna arrays.

\subsubsection{Prototyping Practical ER-FAS Hardware}
Hardware prototypes are fabricated for both the ER-FAS array element and the entire array. The element prototype validates software-controlled fluidic manipulation and radiation pattern reconfigurability, while the array prototype verifies the beampattern and gain enhancement of the complete ER-FAS.

\subsubsection{Anechoic Chamber Measurements and Indoor Communication Trials}
The performance of the ER-FAS prototypes is measured in an anechoic chamber, showing good agreement with simulations. Additionally, real-world communication performance is validated using a software-defined radio (SDR) setup in indoor environments. The trials assess received power levels and bit error rate (BER), confirming the effectiveness of ER-FAS in enhancing practical communication systems.

The paper is organized as follows. Section~\ref{sec:design} introduces the ER-FAS design. Section~\ref{sec:channel modeling} models the wireless channel in both far-field and near-field. A greedy beamforming algorithm is proposed in Section~\ref{sec:beamforming algorithm}. Section~\ref{sec:Full-wave simulation} provides theoretical analysis with full-wave validation. Hardware prototypes and measurement results are presented in Sections~\ref{sec:Hardware} and~\ref{sec:Experiments}. Section~\ref{sec:Conclusion} concludes the paper.

\section{Electromagnetically Reconfigurable Fluid Antenna Design}\label{sec:design}

\subsection{ER-FAS Concept and Array Element Design}\label{subsec:element}

\begin{figure}[t]
  \centering
  \includegraphics[width=\columnwidth]{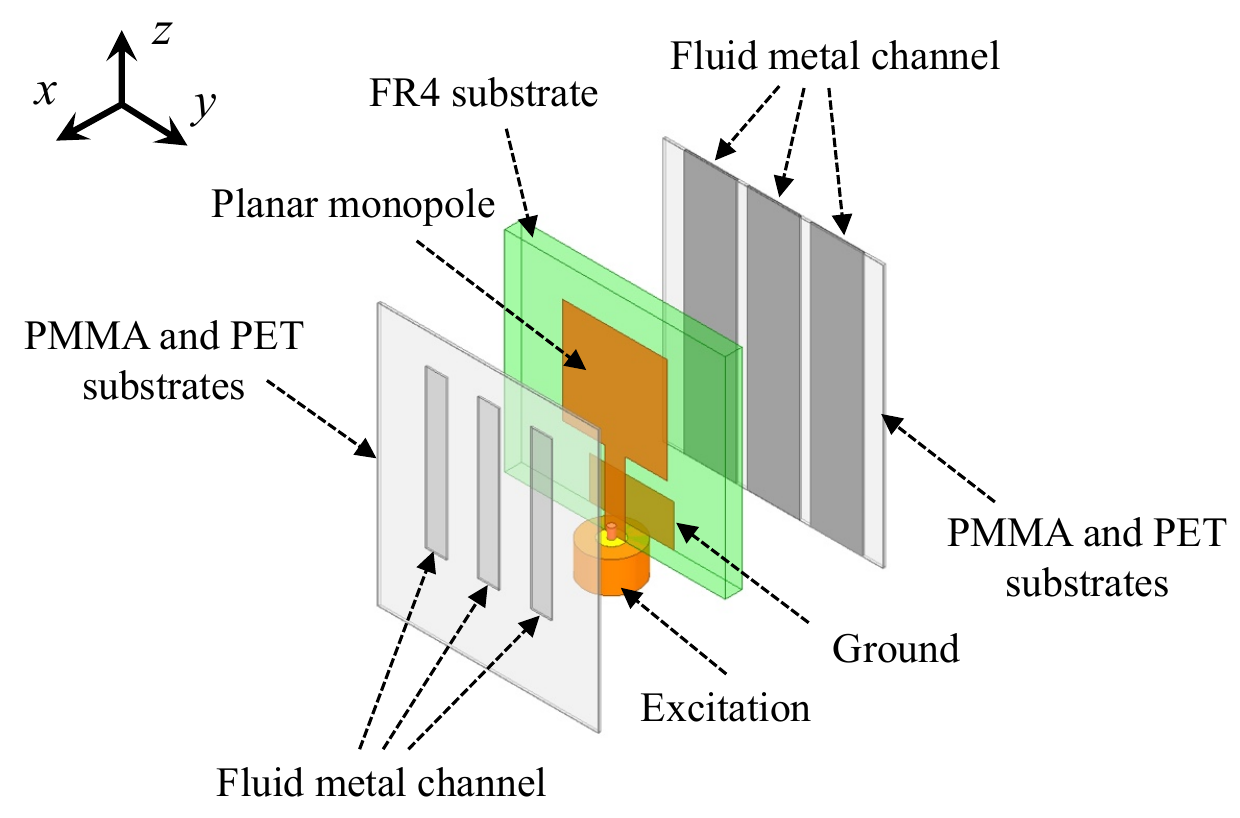}
  \vspace{-1em}
  \caption{
    The proposed ER-FAS array element design.
  }
  \label{fig_FAS_Array_Element}
\end{figure} 

Compared with traditional antenna arrays and SR-FAS, the key feature of the proposed ER-FAS concept is the reconfigurability of the electromagnetic properties of each array element. Based on this concept, a novel practical ER-FAS element design is proposed in Fig.~\ref{fig_FAS_Array_Element}. The proposed ER-FAS array element consists of three horizontally stacked layers. The front layer comprises parasitic fluid metal channels, where the fluid metal is encapsulated within transparent polymethyl methacrylate (PMMA) and polyethylene terephthalate (PET) substrates. The fluid metal can be injected into and extracted from these channels, which are designed to be transparent to facilitate the visualization of the fluid channel states. The injection and extraction of each fluid metal channel can be independently controlled to prevent coupling effects between different reconfigurable states. The middle layer incorporates a planar monopole antenna with two metallic layers. The monopole antenna is excited via an SMA connector. The back layer consists of an additional set of parasitic fluid metal channels, which are also independently controlled and reconfigurable, further enhancing its adaptability.

The detailed operating principle of the proposed fluid antenna element design is described as follows. First, a planar monopole antenna is designed and optimized to achieve wideband impedance performance. The input RF signal is directly fed to the central planar monopole antenna, serving as the excitation source for the entire antenna element. However, the radiation pattern of the designed planar monopole is quasi-omnidirectional, meaning it radiates EM signals across the entire horizontal plane, which is typically undesirable for phased arrays. In contrast, an array element with a directional pattern is preferred for effective antenna array beamforming. Furthermore, with only the planar monopole element, the array remains static and non-reconfigurable.

\begin{figure*}[t]
  \centering
  \includegraphics[width=\linewidth]{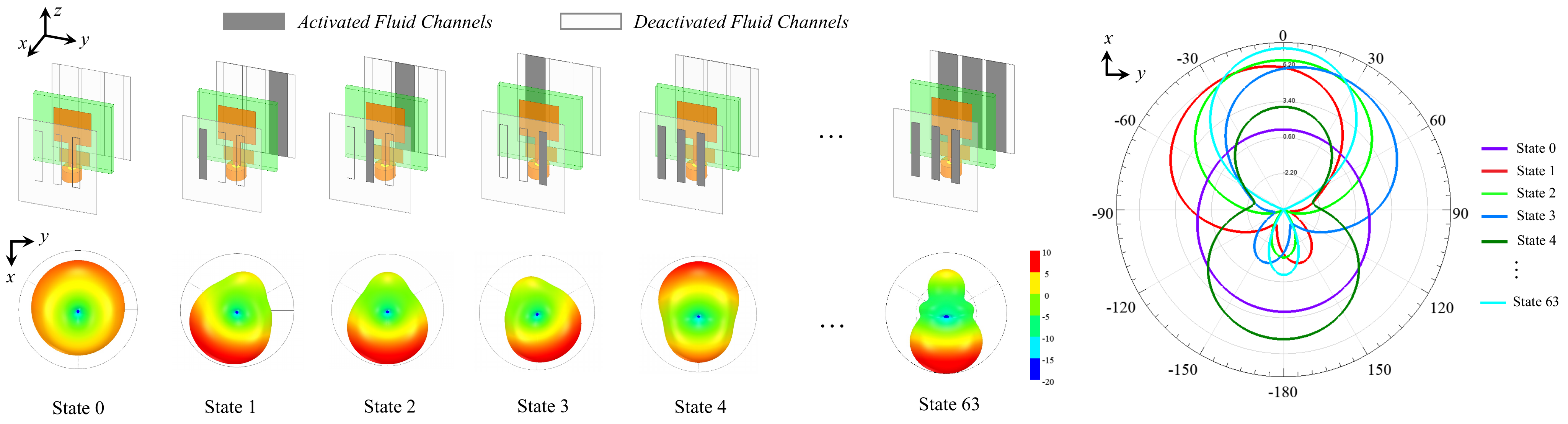}
  \vspace{-1.8em}
  \caption{ 
     The operation principle of the proposed ER-FAS array element and radiation patterns at different reconfigurable states.
    }
  \label{fig_FAS_Element_Mechanism}
\end{figure*}

Inspired by the Yagi-Uda antenna design~\cite{Yagi1928}, we further introduce two parasitic fluid-metal coupling channels into the planar monopole. The front fluid-metal layer serves as a set of directors that guide the EM wave to radiate into the front space, while the back fluid-metal layer acts as a set of reflectors, reducing back-radiation leakage. By incorporating these two fluid-metal layers, the directivity and front-to-back ratio (FBR) of the antenna element can be substantially improved. More importantly, since each fluid-metal channel for both directors and reflectors can be independently controlled, the overall antenna radiation pattern can be reconfigured accordingly. It should be noted that the propagation mode of the electromagnetic signal remains unchanged. Specifically, the central monopole element excites the same radiation mode, while the beam direction and shape are dynamically manipulated through directors and reflectors. To provide a better understanding of the working mechanism of the designed ER-FAS element, a practical element model with various reconfigurable states is demonstrated in Fig.~\ref{fig_FAS_Element_Mechanism}. By injecting or extracting fluid metal in both the front and back channels, the overall antenna radiation pattern can be dynamically adjusted. Here, each channel has two states, filled and empty. Thus, both the director and reflector fluid channels contain three-bit information, leading to an array element with a total of six encoded bits, allowing the designed antenna to achieve 64 ($2^{6}$) reconfigurable states. Note that in this work, we introduce only three independent fluid-metal channels for both director and reflector layers as proof-of-concept. However, additional channels can be incorporated, and utilizing $N$ channels per layer can result in \textbf{$2^{2N}$} reconfigurable states.

The designed ER-FAS element is simulated in ANSYS High-Frequency Structure Simulator (HFSS), and the simulated three-dimensional (3D) radiation patterns are illustrated, as demonstrated in Fig.~\ref{fig_FAS_Element_Mechanism}. Here, for the sake of conciseness, six reconfigurable states (0, 1, 2, 3, 4 and 63) are presented, while the remaining states can be obtained in a similar manner. It can be observed that by configuring the fluid-metal channels, the antenna radiation pattern can be digitally controlled by encoding fluid-metal states. Diverse radiation patterns at different reconfigurable states in the $XOY$-plane are obtained, including radiation pattern direction, FBR, beamwidth, and other parameters. For instance, the radiation pattern can be reconfigured from omnidirectional (state 0) to forward radiation (state 2) and back radiation (state4). The main beam directions can be changed (state 1, 2, and 3). Additionally, the beamwidth can be controlled, as seen in states 2 and 63. Thus, the overall radiation pattern properties of the ER-FAS element can be flexibly optimized and controlled by the parasitic fluid-metal director and reflector channels. Here, it should be highlighted that the proposed ER-FAS fundamentally differs from rotatable antennas~\cite{zheng2025rotatableantenna}, which are categorized as movable antennas~\cite{Zhu_lipeng2024Magzine_MA}. While both aim to achieve radiation pattern reconfigurability, rotatable antennas rely on mechanical rotation to introduce spatial degrees of freedom without altering the intrinsic EM properties of the element itself. In contrast, the ER-FAS does not involve position or orientation changes but enables reconfigurability by dynamically tuning the EM characteristics of each element through fluidic shape control. As a result, rotatable antennas rotate in space with fixed pattern shape scannig, whereas the ER-FAS supports diverse, reconfigurable radiation patterns in EM domain, as illustrated in Fig.~\ref{fig_FAS_Element_Mechanism}.

\subsection{Array Configuration of the proposed ER-FAS}\label{subsec:array}

\begin{figure}[t]
  \centering
  \includegraphics[width=\linewidth]{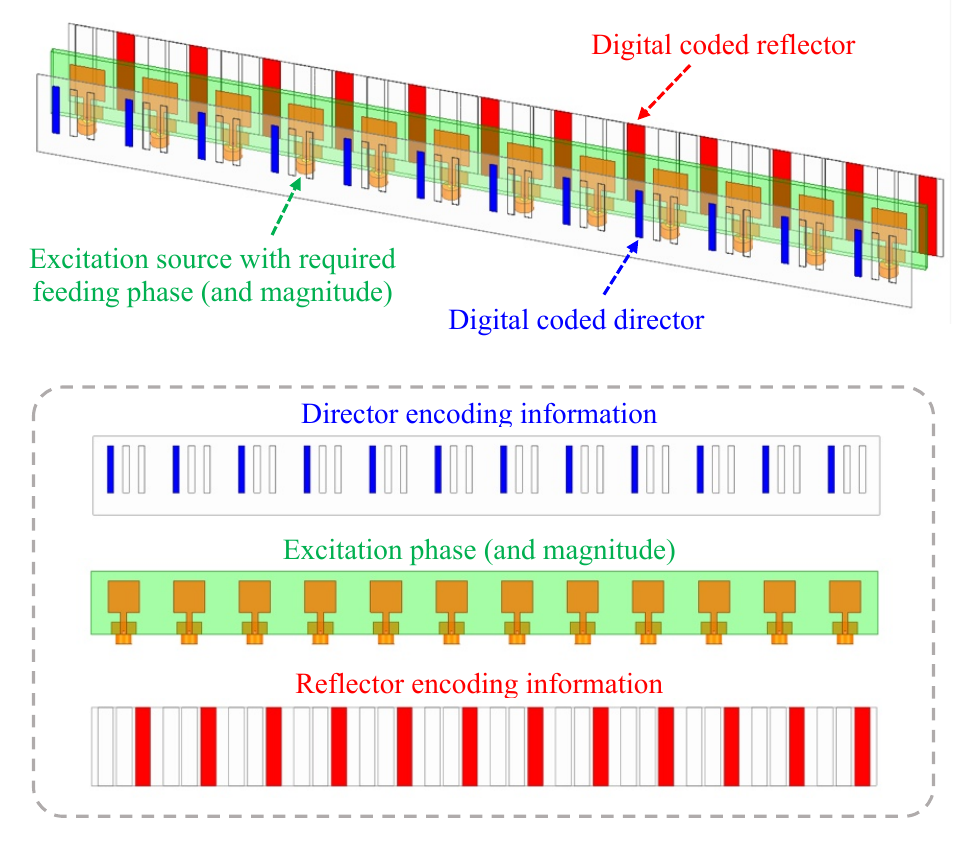}
  \vspace{-2em}
  \caption{ 
     Array configuration of the proposed ER-FAS. The feeding network in this work maintains a fixed excitation magnitude while reconfiguring only the phase, though adjusting both is also achievable.
    }
  \label{fig_FAS_Array}
\end{figure}

Based on the proposed reconfigurable fluid antenna element design, the entire ER-FAS can be constructed. In general, a two-dimensional (2D) array configuration can be constructed. In this work, we investigate a one-dimensional (1D) array configuration for proof of the proposed concept, which can be easily extended to 2D architectures. The designed 1D array architecture with 12 reconfigurable elements is illustrated in Fig.~\ref{fig_FAS_Array}. All the ER-FAS elements operate at \unit[3.55]{GHz}, which is the center frequency of the fifth generation (5G) n78 band. To minimize coupling interference, the array elements are physically spaced at half-wavelength ($\lambda/2$) at the center frequency. This inter-element spacing provides a good trade-off between suppressing mutual coupling and avoiding the emergence of grating lobes.

For the proposed ER-FAS, there are three digitally encoded physical layers. The front layer with directors and the back layer with reflectors can have various combinations of fluid metal presence or absence. The middle layer with the planar monopole array can have different feeding magnitudes and phases. As a result, the ER-FAS hardware provides extensive degrees of freedom to adjust array beampattern and reshape the wireless channel, benefiting applications including but not limited to near-field beam focusing and far-field large-angle beam scanning. In this work, the excitation phase of the planar monopole array is adjustable while the magnitude is maintained. However, it is also feasible to adjust the magnitude to increase the design freedom in practice. Accordingly, the feeding network is designed comprising a 1-to-12 power divider cascaded with 12 independent phase shifters via RF adapters. The power divider provides 12 outputs with equal phases, and the specific phase states for each planar monopole antenna are achieved by the phase shifters that are realized using microstrip lines with different electrical lengths. Detailed simulation results of the designed ER-FAS are presented in Section~\ref{sec:Full-wave simulation}.

\section{Communication System Modeling}\label{sec:channel modeling}
This section models the baseband signal and wireless channel based on the \ac{ER-FAS} proposed in Section~\ref{sec:design}.

\subsection{Signal Model}

\begin{figure*}[t]
  \centering
  \includegraphics[width=0.95\linewidth]{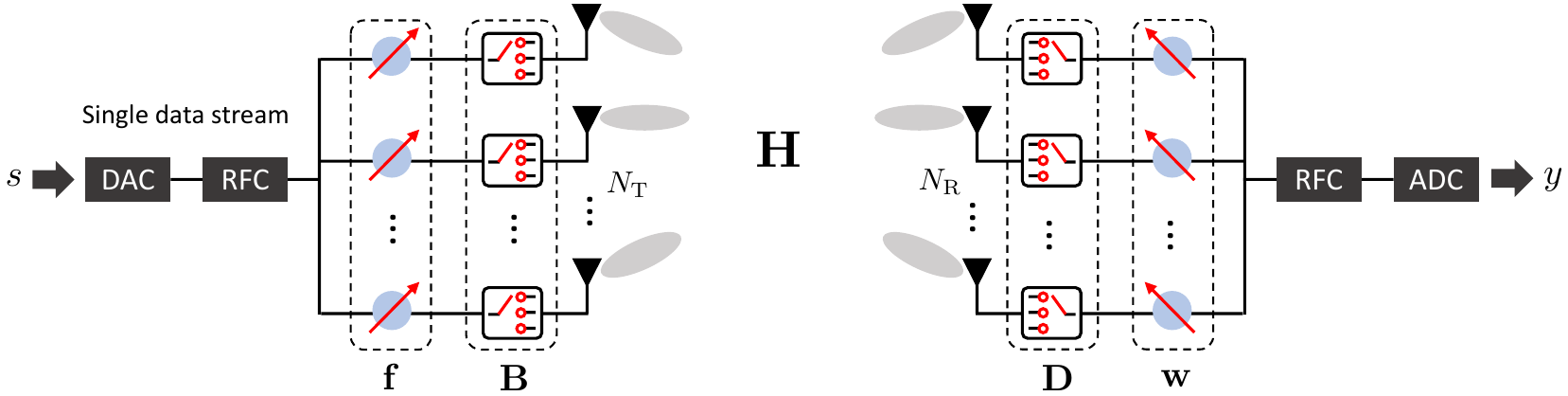}
  \vspace{-0.7em}
  \caption{ 
     Simplified hardware block diagram of a single-user \ac{MIMO} system based on \ac{ER-FAS} and \ac{RF} phase shifters.
    }
  \label{fig_MIMOsystem}
\end{figure*}

We consider a single-user communication system based on \ac{ER-FAS} as demonstrated in Fig.~\ref{fig_MIMOsystem}, where a $N_\Tt$-antenna transmitter (Tx) communicates a single data stream to a $N_\Rt$-antenna receiver (Rx). Both the transmitter and the receiver are equipped with a single \ac{RF} chain. Let~$s\in\mathbb{C}$ denote the transmit symbol. We assume $|s|^2=P_\mathrm{T}$, where $P_\Tt$ represents the transmit power. The transmitter first up-converts the symbol to the carrier frequency by passing through the \ac{RF} chain and then applying an \ac{RF} precoder~$\fv\in\mathbb{C}^{N_\Tt\times 1}$. This \ac{RF} precoder is implemented using analog phase shifters (i.e., the feeding network) with constraint~$|f_i|^2 = 1/N_\Tt$, $i=1,2,\dots,N_\Tt$. After the \ac{RF} precoder, the signal is radiated out via the~$N_\Tt$ transmit antennas, passes through the wireless channel~$\Hm\in\mathbb{C}^{N_\Rt\times N_\Tt}$, and is received by the~$N_\Rt$ receive antennas. Note that in our proposed antenna design, various radiation patterns can be chosen for each antenna individually. This reconfigurability will be reflected in the expression of channel matrix~$\Hm$ later.

The receiver applies a \ac{RF} combiner~$\wv\in\mathbb{C}^{N_\Rt\times 1}$ and then down-converts the radio signal to the baseband through the \ac{RF} chain. Similarly, we constrain the combiner as $|w_j|^2=1/N_\Rt$, $j=1,2,\dots,N_\Rt$. Hence, the final received baseband signal is given by \begin{equation}\label{eq:BBsignal}
    y = \wv^\HH\Hm\fv s + \wv^\HH\nv,
\end{equation}
where~$\nv\sim\mathcal{CN}(\mathbf{0},\sigma^2\mathbf{I}_{N_\Rt})$ denotes additive white Gaussian noise. Therefore, the overall spectral efficiency (rate) of such a system assuming Gaussian signaling is given by~\cite{Goldsmith2003Capacity,Ayach2014Spatially,Sohrabi2016Hybrid}
\begin{equation}\label{eq:Rate}
    R = \log_2 \Big(1+\frac{P_\Tt|\wv^{\HH}\Hm\fv|^2}{\sigma^2}\Big).
\end{equation}
In the following subsections, we model the wireless channel~$\Hm$ in both the far-field and near-field scenarios.

\subsection{Far-Field Channel Model}
In this subsection, we express the channel matrix~$\Hm$ in the far-field condition, which is characterized by the \emph{planar wave model}. We will show that our proposed reconfigurable antenna design introduces extra degrees of freedom to reshape the wireless channel. We start by analyzing the wireless channel based on the conventional antenna array and then extend to the proposed \ac{ER-FAS}. 

\subsubsection{Conventional Antenna Array}
We consider a scenario where a \ac{LoS} and multiple \ac{NLoS} paths exist. Based on the \ac{SV} multipath model, the \ac{NLoS} channel is modeled as the sum of the contributions of $C$ scattering clusters, with the $c^\text{th}$ cluster contributing $L_c$ propagation paths. Assuming an ideal Dirac pulse-shaping filter, the equivalent frequency-domain baseband channel based on a conventional antenna array can be expressed as~\cite{Ayach2014Spatially,Tarboush2021TeraMIMO}
\begin{multline}\label{eq:HFF_conv}
    \Hm_\mathrm{CV} = \gamma\alpha_{\mathrm{LoS}}G_\Rt(\phiv_\mathrm{LoS})G_\Tt(\thetav_\mathrm{LoS})\av_\Rt(\phiv_\mathrm{LoS})\av_\Tt^\HH(\thetav_\mathrm{LoS}) \\
    +\gamma\sum_{c=1}^C\sum_{\ell=1}^{L_c}\alpha_{c,\ell}G_\Rt(\phiv_{c,\ell})G_\Tt(\thetav_{c,\ell})\av_\Rt(\phiv_{c,\ell})\av_\Tt^\HH(\thetav_{c,\ell}),
\end{multline}
where~$\gamma$ is a normalization factor such that~$\gamma = \sqrt{\frac{N_\Tt N_\Rt}{1+\sum_{c=1}^C L_c}}$. Here, $\alpha$ denotes complex channel gain, $\thetav$ denotes \ac{AoD} at the transmitter, and $\phiv$ denotes \ac{AoA} at the receiver, corresponding to different paths. Note that each angle contains an azimuth and elevation components, i.e., $\thetav = [\theta_{\mathrm{az}},\theta_{\mathrm{el}}]^\TT$, $\phiv = [\phi_{\mathrm{az}},\phi_{\mathrm{el}}]^\TT$. For clarity, the azimuth angle is defined as the angle between the positive $X$-axis and the target's projection on the $XOY$-plane, while the elevation angle is the angle between the $Z$-axis and the target direction. Both are expressed in the transmitter's and receiver's body frames. In addition, $G_\Tt$ and $G_\Rt$ represent the transmit and receive \emph{antenna gains},\footnote{The antenna gains $G_\Tt$ and $G_\Rt$ used in this paper refer to the \emph{magnitude} gain of antennas, as defined in~\cite[Eq.~(7)]{Zheng2025Trihybrid}. Specifically, the antenna gain is defined as the radiation intensity of an antenna in a given direction relative to that
of an isotropic radiator~\cite{Balanis2016Antenna}; thus it is inherently normalized and satisfies the constraint $\int_{0}^{2\pi}\int_0^\pi G_\Tt^2(\thetav)\sin{\theta_\mathrm{el}} \mathrm{d}\theta_\mathrm{el} \mathrm{d}\theta_\mathrm{az}=4\pi$. A detailed elaboration and derivation can be found in~\cite[Sec.~II-B]{Zheng2025Trihybrid}.} which are functions of corresponding \ac{AoD} and \ac{AoA}. Finally, the vectors $\av_\Tt$ and $\av_\Rt$ are the normalized receive and transmit \acp{ARV}. Taking $\av_\Tt$ as an example, assuming an $N_\Tt^{\mathrm{h}}\times N_\Tt^{\mathrm{v}}$ \ac{UPA} configuration (thus $N_\Tt=N_\Tt^{\mathrm{h}} N_\Tt^{\mathrm{v}}$), the transmit \ac{ARV} is written as
\begin{equation}
    \av_\Tt(\thetav) = \frac{1}{\sqrt{N_\Tt}}e^{-j2\pi\theta^{\mathrm{h}}\kv(N_\Tt^\mathrm{h})}\!\otimes\! e^{-j2\pi\theta^\mathrm{v}\kv(N_\Tt^\mathrm{v})},
\end{equation}
where~$\kv(N)=[0,1,\dots,N-1]^\TT$, and~$\theta^\mathrm{h}$ and~$\theta^\mathrm{v}$ are the spatial angles corresponding to the horizontal and vertical dimensions, respectively. Assuming the \ac{UPA} is deployed on the $YOZ$-plane of the transmitter's body coordinate system, we obtain $\theta^\mathrm{h}\triangleq{d_\It}\sin(\theta_\mathrm{az})\sin(\theta_\mathrm{el})/{\lambda}$ and $\theta^\mathrm{v}\triangleq{d_\It}\cos(\theta_\mathrm{el})/{\lambda}$, where~$\lambda$ is the wavelength of the operating frequency and $d_\It$ is the inter-element spacing of the transmit antenna array. The receive \ac{ARV} $\av_\Rt(\phiv)$ is defined in the same way.

\subsubsection{Electromagnetically Reconfigurable Fluid Antenna Array}
In conventional antenna arrays, it is trivial to see that each antenna element shares the same radiation pattern, which is characterized by antenna gain function~$G_\Tt(\thetav)$/$G_\Rt(\phiv)$. 
When using the element-reconfigurable array, these antenna elements can choose different radiation patterns by adjusting the state of the liquid conducting material as demonstrated in Section~\ref{sec:design}. We use~$G_{\Tt,i}(\thetav)$ and $G_{\Rt,j}(\phiv)$ to denote the radiation patterns of the~$i^\text{th}$ transmit antenna and the~$j^\text{th}$ receive antenna, respectively. Each of these radiation patterns is chosen from a preset set of $N$ available radiation patterns, which we denote as~$\{\bar{G}_1,\bar{G}_2,\dots,\bar{G}_N\}$. Defining a dictionary vector $\bar{\gv}(\varphiv) \triangleq [\bar{G}_1(\varphiv),\bar{G}_2(\varphiv),\dots,\bar{G}_N(\varphiv)]^\TT$, we can write
\begin{align}
    G_{\Tt,i}(\thetav) &= \bar{\gv}(\thetav)^\TT \bv_{\Tt,i}, \quad i=1,2,\dots, N_\Tt,\\
    G_{\Rt,j}(\phiv) &= \bar{\gv}(\phiv)^\TT \bv_{\Rt,j}, \quad j=1,2,\dots, N_\Rt.
\end{align}
Here, $\bv_{\Tt,i}$ and $\bv_{\Rt,j}$ are two binary vectors denoting the selection of radiation pattern, which are constrained as 
\begin{equation}\label{eq:ConsSelection}
    \bv_{\Tt,i},\bv_{\Rt,j}\in\{\bv|\bv\in\{0,1\}^N,\ \|\bv\|_2=1\}.
\end{equation}
For notational convenience, we assume that the transmit and receive antennas share the same set of reconfigurable radiation patterns.

Based on this radiation pattern selection mechanism, we can extend~\eqref{eq:HFF_conv} to our \ac{ER-FAS} as
\begin{align}\label{eq:HFF_new1}
    &\hspace{-0.5em}\Hm_\mathrm{ER}\! =\! \gamma\alpha_\mathrm{LoS}\big(\gv_\Rt(\phiv_\mathrm{LoS})\!\odot\!\av_\Rt(\phiv_\mathrm{LoS})\big)  \big(\gv_\Tt(\thetav_\mathrm{LoS})\!\odot\!\av_\Tt(\thetav_\mathrm{LoS})\big)^\HH\notag\\
    &\hspace{-0.5em}+ \!\gamma\!\sum_{c=1}^C\!\sum_{\ell=1}^{L_c}\!\!\alpha_{c,\ell}\!\big(\gv_\Rt(\phiv_{c,\ell})\!\odot\!\av_\Rt(\phiv_{c,\ell})\big) \!   \big(\gv_\Tt(\thetav_{c,\ell})\!\odot\!\av_\Tt(\thetav_{c,\ell})\big)^\HH\!\!\!,\!
\end{align}
where $\odot$ denotes the Hadamard product, and $\gv_\Xt(\varphiv)=[G_{\Xt,1}(\varphiv),G_{\Xt,2}(\varphiv),\dots,G_{\Xt,N_{\Xt}}(\varphiv)]^\TT$ for $\Xt\in\{\Tt,\Rt\}$. We further define the following two selection matrices $\Bm$ and $\Dm$ for the transmitter and receiver, respectively:
\begin{align}
    \Bm &= \text{blkdiag}\{\bv_{\Tt,1}^\TT,\bv_{\Tt,2}^\TT,\dots,\bv_{\Tt,N_\Tt}^\TT\}\in\mathbb{R}^{N_\Tt\times NN_\Tt},\label{eq:Bdef}\\
    \Dm &= \text{blkdiag}\{\bv_{\Rt,1}^\TT,\bv_{\Rt,2}^\TT,\dots,\bv_{\Rt,N_\Rt}^\TT\}\in\mathbb{R}^{N_\Rt\times NN_\Rt}.\label{eq:Ddef}
\end{align}
Then, we have
\begin{align}  \gv_\Tt(\thetav)\odot\av_\Tt(\thetav)&=\Bm\big(\av_\Tt(\thetav)\otimes\bar{\gv}(\thetav)\big),\label{eq:gtat}\\
\gv_\Rt(\phiv)\odot\av_\Rt(\phiv)&=\Dm\big(\av_\Rt(\phiv)\otimes\bar{\gv}(\phiv)\big),\label{eq:grar}
\end{align}
where $\otimes$ denotes the Kronecker product. 
Substituting \eqref{eq:gtat} and~\eqref{eq:grar} into \eqref{eq:HFF_new1}, we obtain
\begin{equation}\label{eq:HERFF_2}
    \Hm_\mathrm{ER} = \gamma\Dm{\Hm}_\mathrm{EM}\Bm^\TT,
\end{equation}
where~${\Hm}_\mathrm{EM}\in\mathbb{C}^{NN_\Rt\times NN_\Tt}$ is called the \ac{EM}-domain channel~\cite{Ying2024Reconfigurable} given by
\begin{multline}\label{eq:tildeHER}
    {\Hm}_\mathrm{EM} = \alpha_\mathrm{LoS}\big(\av_\Rt(\phiv_\mathrm{LoS})\!\otimes\!\bar{\gv}(\phiv_\mathrm{LoS})\big)\big(\av_\Tt(\thetav_\mathrm{LoS})\!\otimes\!\bar{\gv}(\thetav_\mathrm{LoS})\big)^\HH\\
    \!+ \!\sum_{c=1}^C\!\sum_{\ell=1}^{L_c}\!\alpha_{c,\ell}\big(\av_\Rt(\phiv_{c,\ell})\!\otimes\!\bar{\gv}(\phiv_{c,\ell})\big)  \big(\av_\Tt(\thetav_{c,\ell})\!\otimes\!\bar{\gv}(\thetav_{c,\ell})\big)^\HH\!\!\!.
\end{multline}
The matrices $\Bm$ and $\Dm$ fully describe the radiation pattern configurations of all antennas at the transmitter and receiver, as depicted in Fig.~\ref{fig_MIMOsystem}.

\subsection{Near-Field Channel Model}

We now extend the proposed \ac{ER-FAS} channel model to near-field scenarios, where the wireless channel is described by the \emph{spherical wave model}~\cite{Cui2023Near,Tarboush2024Cross}. For a near-field receiver, the signal transmitted from each antenna exhibits distinct amplitude, phase, and \ac{AoD}/\ac{AoA}. Note that under the far-field model, the signals associated with different antenna elements in an array share the same amplitude and \ac{AoD}/\ac{AoA}, differing only in phase. To characterize the spherical wave model, we use $\pv_{\mathrm{T},i}$ and $\pv_{\mathrm{R},j}$ to denote the absolute positions of the $i^\text{th}$ antenna at the transmitter and the $j^\text{th}$ antenna at the receiver, respectively. Additionally, the position of the ${\ell}^\text{th}$ scatterer in the $c^\text{th}$ scattering cluster is denoted as $\rv_{c,\ell}$. Then, the near-field version of~\eqref{eq:tildeHER} is given by~\cite{Liu2023Near}
\begin{equation}\label{eq:HEMNF}
    {\Hm}_\mathrm{EM} = {\Hm}_\mathrm{EM}^\mathrm{LoS} + \sum_{c=1}^C\sum_{\ell=1}^{L_c} \tilde{\av}_\Rt(\rv_{c,\ell})  \tilde{\av}_\Tt(\rv_{c,\ell})^\HH.
\end{equation}

Under the spherical wave model, the \ac{EM}-domain \ac{LoS} channel  ${\Hm}_\mathrm{EM}^\mathrm{LoS}\in\mathbb{C}^{NN_\Rt\times NN_\Tt}$ can be partitioned as 
\begin{equation}
  {\Hm}_\mathrm{EM}^\mathrm{LoS} = \begin{bmatrix}
      {\Hm}_\mathrm{EM}^\mathrm{LoS}(1,1) & \cdots & {\Hm}_\mathrm{EM}^\mathrm{LoS}(N_\Tt,1) \\
      \vdots & \ddots & \vdots \\
      {\Hm}_\mathrm{EM}^\mathrm{LoS}(1,N_\Rt) & \cdots & {\Hm}_\mathrm{EM}^\mathrm{LoS}(N_\Tt,N_\Rt)
  \end{bmatrix},  
\end{equation}
where 
\begin{equation}
    {\Hm}_\mathrm{EM}^\mathrm{LoS}(i,j) = \alpha_{i,j}e^{-j\frac{2\pi}{\lambda}\|\pv_{\Tt,i}-\pv_{\Rt,j}\|_2}\bar{\gv}(\phiv_{i,j})\bar{\gv}(\thetav_{i,j})^\TT\in\mathbb{C}^{N\times N}.\notag
\end{equation}
Here, $\alpha_{i,j}$, $\thetav_{i,j}$, and $\phiv_{i,j}$ respectively denote the distinct channel gain, \ac{AoD}, and \ac{AoA} of the \ac{LoS} link from the $i^\text{th}$~antenna at the transmitter to the $j^\text{th}$~antenna at the receiver. In addition, the \acp{ARV} in~\eqref{eq:HEMNF} can be partitioned as
\begin{align}
    \tilde{\av}_\Tt(\rv_{c,\ell}) &= [\tilde{\av}_{\Tt,1}(\rv_{c,\ell})^\TT,\dots,\tilde{\av}_{\Tt,N_\Tt}(\rv_{c,\ell})^\TT]^\TT\in\mathbb{C}^{NN_\Tt}, \\  \tilde{\av}_\Rt(\rv_{c,\ell}) &= [\tilde{\av}_{\Rt,1}(\rv_{c,\ell})^\TT,\dots,\tilde{\av}_{\Rt,N_\Rt}(\rv_{c,\ell})^\TT]^\TT \in\mathbb{C}^{NN_\Rt},
\end{align}
where
\begin{align}
    \tilde{\av}_{\Tt,i}(\rv_{c,\ell}) &= \alpha_{i,c,\ell}e^{-j\frac{2\pi}{\lambda}\|\pv_{\Tt,i}-\rv_{c,\ell}\|_2}\bar{\gv}(\thetav_{i,c,\ell})\in\mathbb{C}^{N},\\
    \tilde{\av}_{\Rt,j}(\rv_{c,\ell}) &= \beta_{j,c,\ell}e^{-j\frac{2\pi}{\lambda}\|\pv_{\Rt,j}-\rv_{c,\ell}\|_2}\bar{\gv}(\phiv_{j,c,\ell})\in\mathbb{C}^{N}.
\end{align}
Here, $\alpha_{i,c,\ell}$ and $\thetav_{i,c,\ell}$ respectively denote the channel gain and \ac{AoD} of the link from the $i^\text{th}$ antenna at the transmitter to the $\ell^\text{th}$ scatterer in the $c^\text{th}$ scattering cluster. Similarly, $\beta_{j,c,\ell}$ and $\phiv_{j,c,\ell}$ respectively denote the channel gain and \ac{AoA} of the link from the $\ell^\text{th}$ scatterer in the $c^\text{th}$ scattering cluster to the $j^\text{th}$ antenna at the receiver.

\section{Joint Beamforming Design}\label{sec:beamforming algorithm}
Based on~\eqref{eq:Rate} and~\eqref{eq:HERFF_2}, we can formulate the following joint beamforming problem
\begin{align}\label{eq:opt}
    \max_{\fv,\wv,\Bm,\Dm}&\quad \log_2 \Big(1+\frac{\gamma P_\Tt|\wv^{\HH}\Dm{\Hm}_\mathrm{EM}\Bm^\TT\fv |^2}{\sigma^2}\Big),\\
    \text{s.t.}&\quad |f_i|^2 = \frac{1}{N_\Tt},\ |w_j|^2=\frac{1}{N_\Rt},~\eqref{eq:ConsSelection},~\eqref{eq:Bdef},~\eqref{eq:Ddef},\notag
\end{align}
where $\Hm_\mathrm{EM}$ can be either a far-field channel (as defined in~\eqref{eq:tildeHER}) or a near-field channel (as defined in~\eqref{eq:HEMNF}) in the \ac{EM} domain. Here, $\fv$ and $\wv$ represent the analog phase shifters, while $\Bm$ and $\Dm$ characterize the radiation pattern selection of these antenna elements. Supposing the electromagnetic-domain channel $\Hm_\mathrm{EM}$ is known,\footnote{While many channel estimation techniques can be used to estimate $\Hm_\mathrm{EM}$ from received pilot signals~\cite{Alkhateeb2014Channel,Liu2018Massive}, estimation errors are inevitable, making perfect channel state information (CSI) generally unrealistic in practice. However, as an initial proof-of-concept study, we adopt this assumption to avoid unnecessary complexity. For the same reason, we also assume that both the transmitter and receiver have perfect knowledge of the available antenna radiation patterns at the opposite end. Relaxing these assumptions and developing robust techniques under imperfect channel and pattern knowledge represent important directions for future work.} this subsection proposes a low-complexity solution to~\eqref{eq:opt}. 

\subsection{An Alternating Beamforming Algorithm}

The formulated joint beamforming problem can be solved by optimizing $\{\fv,\wv,\Bm,\Dm\}$ alternately until they converge, i.e., applying the \emph{block-coordinate descent method}~\cite{Beck2013On}. The optimization of the analog phase shifters is straightforward given the other variables. For example, when updating $\fv$ given $\{\wv,\Bm,\Dm\}$, we have $\fv^\star = \frac{1}{\sqrt{N_\Tt}}e^{j\arg(\Bm{\Hm}_\mathrm{EM}^\HH\Dm^\TT\wv)}$. The same principle can be applied to optimizing $\wv$ given $\{\fv,\Bm,\Dm\}$. The main challenge lies in optimizing the antenna state selection matrices $\Bm$ and $\Dm$. The globally optimal solutions for $\Bm$ and $\Dm$ can be determined through exhaustive search; however, this approach becomes impractical due to the exponentially increasing computational complexity as the size of the antenna arrays grows. To address this problem, we propose a greedy approach to obtain a suboptimal solution for $\Bm$ and $\Dm$, with computational complexity scaling quadratically (instead of exponentially) with the number of antennas.

Let's focus on optimizing $\Bm$ given $\{\fv,\wv,\Dm\}$. This subproblem is equivalent to
\begin{align}\label{eq:optB}
    \max_{\Bm}\quad &\big|\wv^\HH\Dm\Hm_\mathrm{EM}\Bm^\TT\fv\big|,\\
    \text{s.t.}\quad &\eqref{eq:ConsSelection},\  \eqref{eq:Bdef}.\notag
\end{align}
According to \eqref{eq:ConsSelection} and  \eqref{eq:Bdef}, we can rewrite the objective here as $\big|\dv^\TT[\bv_{\Tt,1}^\TT,\bv_{\Tt,2}^\TT,\dots,\bv_{\Tt,N_\Tt}^\TT]^\TT\big|$, where $\dv = (\Hm_\mathrm{EM}^\TT\Dm^\TT\wv^*)\odot(\fv\otimes\mathbf{1}_N)\in\mathbb{C}^{NN_\Tt}$. We further denote $\dv_m=[\dv]_{(m-1)N+1:mN}\in\mathbb{C}^{N},\ m=1,2,\dots,N_\Tt$. Then,~\eqref{eq:optB} can be rewritten as 
\begin{align}\label{eq:optB2}
    \max_{\{\bv_{\Tt,m}\}_{m=1}^{N_\Tt}}\quad &\Big|\sum_{m=1}^{N_\Tt}\dv_m^\TT\bv_{\Tt,m}\Big|,\\
    \text{s.t.}\quad &\eqref{eq:ConsSelection}.\notag
\end{align}
This implies that the objective is the sum of the contributions from one entry selected from each $\dv_m,\ m=1,2,\dots,N_\Tt$. Therefore, our proposed strategy is as follows: we first sum all the elements in all $\dv_m$ to compute a complex objective value $v=\sum_{m=1}^{N_t}\dv_m^\TT\mathbf{1}_N$. Next, we remove one entry with the smallest contribution to the objective value from each $\dv_m$. That is, the index of the removed entry in $\dv_m$ is determined by
\begin{equation}
u_m = \arg\max_{u}\quad \big|v-[\dv_m]_u\big|.   
\end{equation}
We then update the objective value by summing all the remaining entries and iteratively removing the least contributing element from each $\dv_m$ based on the current objective value until only one entry remains in each $\dv_m$. Finally, the positions of the only remaining entries in these $\dv_m$ are marked as the selected antenna states. That is, the corresponding positions in $\bv_{\mathrm{T},N_\mathrm{T}}$ are set to 1 while all others are set to 0. The optimized matrix $\Bm$ is then obtained according to~\eqref{eq:Bdef}. The optimization of $\Dm$, given ${\fv, \wv, \Bm}$, can be tackled using the same principle. The complete optimization process is summarized in Algorithm~\ref{algo:1}.

 \begin{algorithm}[t]
 \caption{A Greedy Algorithm for Solving~\eqref{eq:opt}}
 \label{algo:1}
 \begin{algorithmic}[1]
 \State \textbf{Input:}  ${\Hm}_\mathrm{EM}$\qquad \textbf{Output:}  $\fv^\star,\wv^\star,\Bm^\star,\Dm^\star$
\State Set $k=0$ and randomly initialize $\fv_0,\wv_0,\Bm_0,\Dm_0$.
\Statex{\hspace{-1.2em}\textbf{Repeat:}}

\Statex{\hspace{-0.9em}\textbf{Optimize $\fv$:}} $\max_\fv\ |\wv_k^\HH\Dm_k{\Hm}_\mathrm{EM}\Bm_k^\TT \fv|$,\ s.t.\ $|f_i|^2=\frac{1}{N_\Tt}$
\State $\fv_{k+1}=\frac{1}{\sqrt{N_\Tt}}e^{j\arg(\Bm_k{\Hm}_\mathrm{EM}^\HH\Dm_k^\TT\wv_k)}$. 

\Statex{\hspace{-0.9em}\textbf{Optimize $\Bm$:}} $\max_{\Bm}\ |\wv_k^\HH\Dm_k{\Hm}_\mathrm{EM}\Bm^\TT \fv_{k+1}|$,\ s.t.~\eqref{eq:ConsSelection}, \eqref{eq:Bdef}
\State \label{step:begin}Compute $\dv_m\!=\!\big[(\Hm_\mathrm{EM}^\TT\Dm_k^\TT\wv_k^*)\odot(\fv_{k+1}\!\otimes\mathbf{1}_N)\big]_{(m\!-\!1)N\!+\!1:mN}.$ 
\State Define $N_\Tt$ sets $\mathcal{S}_m=\{1,2,\dots,N\}$, $m=1,2,\dots,N_\Tt$.
\State Set $[\bv_{\Tt,m}]_{\mathcal{S}_m}=1$, otherwise zeros, for $m=1,2,\dots,N_\Tt$.
\For{$n=1,2,\dots,N-1$}
    \State Compute $v=\sum_{m=1}^{N_t}\dv_m^\TT\bv_{\Tt,m}$.
    \For{$m=1,2,\dots,N_\Tt$}
        \State $u_m = \arg\max_{u\in\mathcal{S}_m}\ \big|v\!-\![\dv_m]_u\big|$, $\mathcal{S}_m = \mathcal{S}_m $\textbackslash$u_m$. 
        \State Reset $[\bv_{\Tt,m}]_{\mathcal{S}_m}=1$, otherwise zeros.
    \EndFor
\EndFor
\State \label{step:end}$\Bm_{k+1} = \text{blkdiag}\{\bv_{\Tt,1}^\TT,\bv_{\Tt,2}^\TT,\dots,\bv_{\Tt,N_\Tt}^\TT\}.$ 

\Statex{\hspace{-0.9em}\textbf{Optimize $\wv$:}} $\max_\wv\!\ |\wv^\HH\Dm_k{\Hm}_\mathrm{EM}\Bm_{k+1}^\TT \fv_{k+1}|$,\ s.t.\!\ $|w_i|^2=\frac{1}{N_\Rt}$
\State $\wv_{k+1}=\frac{1}{\sqrt{N_\Rt}}e^{j\arg(\Dm_k{\Hm}_\mathrm{EM}\Bm_{k+1}^\TT\fv_{k+1})}$. 

\Statex{\hspace{-0.9em}\textbf{Optimize $\Dm$:}} $\max_{\Dm}\ |\wv_{k+1}^\HH\Dm{\Hm}_\mathrm{EM}\Bm_{k+1}^\TT \fv_{k+1}|$,\ s.t.~\eqref{eq:ConsSelection}, \eqref{eq:Ddef}
\State Rewrite the objective function as $|\fv_{k+1}^\HH\Bm_{k+1}\tilde{\Hm}_{\mathrm{ER}}^\HH\Dm^\TT\wv_{k+1}|$ and reapply Step~\ref{step:begin}--\ref{step:end} to obtain $\Dm_{k+1}$.

\State $k=k+1$.

\Statex{\hspace{-1.2em}\textbf{Until}} convergence

\Statex{\hspace{-1.2em}\textbf{Return}} $\fv_k,\wv_k,\Bm_k,\Dm_k$

\end{algorithmic} 
\end{algorithm}

\subsection{Convergence Analysis}

Now we analyze the convergence of Algorithm~\ref{algo:1}. First, it is straightforward to observe that the objective value in~\eqref{eq:opt} is upper bounded. This follows from the fact that the entire variable space is a product of a compact continuous set and a finite discrete set, and the objective function is continuous. A more rigorous justification can be provided using the Cauchy-Schwarz inequality:
\begin{equation}\notag
    \big|\wv^\HH\Dm\Hm_\mathrm{EM}\Bm^\TT\fv\big| \!\leq\! \|\wv\|_2 \cdot \|\Dm\Hm_\mathrm{EM}\Bm^\TT\|_2 \cdot \|\fv\|_2 \!=\! \|\Dm\Hm_\mathrm{EM}\Bm^\TT\|_2,
\end{equation}
where the norm $\|\Dm\Hm_\mathrm{EM}\Bm^\TT\|_2$ must be upper bounded since both $\Bm$ and $\Dm$ lie in finite discrete sets.

Next, we examine the updates performed in the alternating optimization. It is evident that the updates of $\fv$ and $\wv$ in Steps~3 and~13 yield the exact optimum for the corresponding subproblems. Thus, these updates are strictly non-decreasing in the objective value. In contrast, the updates of $\Bm$ and $\Dm$ rely on a greedy selection strategy and may not guarantee strict improvement. To address this, a simple comparison step can be added: the new $\Bm$ and $\Dm$ are adopted only if they result in a higher objective value than the previous candidates. This ensures that all updates are non-decreasing.

As a result, the sequence of objective values generated by Algorithm~\ref{algo:1} is non-decreasing and upper bounded, and hence convergent by the monotone convergence theorem. However, the converged solution is not necessarily a global optimum of~\eqref{eq:opt}. Developing methods to find the global optimum of~\eqref{eq:opt} within reasonable computational complexity remains an important yet challenging direction for future research.

\subsection{Computational Complexity Analysis}\label{sec:complexity}
The proposed Algorithm~\ref{algo:1} offers significantly higher computational efficiency compared to exhaustive search for determining the optimal $\Bm$ and $\Dm$. To demonstrate this advantage, we first analyze the computational complexity of Algorithm~\ref{algo:1}, and then compare it with that of its exhaustive search counterpart. 

In Algorithm~\ref{algo:1}, the optimization is conducted alternately over $\fv$, $\Bm$, $\wv$, and $\Dm$. The corresponding computational complexities are summarized as follows:
\begin{itemize}
    \item The complexity of optimizing $\fv$ in Step~3 is given by $\mathcal{O}(N^2N_\Tt^2N_\Rt+NN_\Tt N_\Rt^2)$.
    \item The complexity of optimizing $\Bm$ in Steps~4--12 is dominated by Step~4, which yields a complexity $\mathcal{O}(N^2N_\Tt N_\Rt^2)$. 
    \item The complexity of optimizing $\wv$ in Step~13 is given by $\mathcal{O}(N^2N_\Tt N_\Rt^2+NN_\Tt^2 N_\Rt)$.
    \item The complexity of optimizing $\Dm$ in Steps~14 is given by $\mathcal{O}(N^2N_\Tt^2 N_\Rt)$. 
\end{itemize}
Assuming the alternating optimization loop is repeated $I_\mathrm{ALT}$ times, the overall computational complexity of Algorithm~\ref{algo:1} is $\mathcal{O}\big(I_\mathrm{ALT}(N^2N_\Tt^2 N_\Rt + N^2N_\Tt N_\Rt^2)\big)$, which scales quadratically with $N_\Tt$, $N_\Rt$, and $N$.

We now consider an exhaustive search-based solution. Although the sets of feasible $\Bm$ and $\Dm$ are finite, the design spaces for the analog beamformers $\fv$ and $\wv$ are continuous and thus infinite. To address this, we adopt an alternating optimization framework similar to Algorithm~\ref{algo:1}, with the only difference being that the optimizations of $\Bm$ and $\Dm$ are replaced by exhaustive searches, while the other steps remain unchanged. Let's first consider optimizing $\Bm$ using exhaustive search. Based on~\eqref{eq:optB2}, $\Bm$ has $N^{N_\Tt}$ possible solutions. For each candidate, the evaluation of objective value in~\eqref{eq:optB2} features a complexity $\mathcal{O}(NN_\Tt)$. Therefore, optimization $\Bm$ using exhaustive search has a complexity $\mathcal{O}(N^{N_\Tt}NN_\Tt)$. Similarly, optimizing $\Dm$ using exhaustive search has a complexity $\mathcal{O}(N^{N_\Rt}NN_\Rt)$. Then, according to the previously derived complexity of optimizing $\fv$ and $\wv$, the overall computational complexity of the exhaustive search solution is $\mathcal{O}\big(I_\mathrm{ALT}(N^2N_\Tt^2N_\Rt + N^{N_\Tt}NN_\Tt + N^2N_\Tt N_\Rt^2 + N^{N_\Rt}NN_\Rt)\big)$. This complexity scales exponentially with the number of antennas $N_\Tt$ and $N_\Rt$.

\section{Simulation Results}\label{sec:Full-wave simulation}

\begin{figure*}[t]
  \centering
  \includegraphics[width=0.9\linewidth]{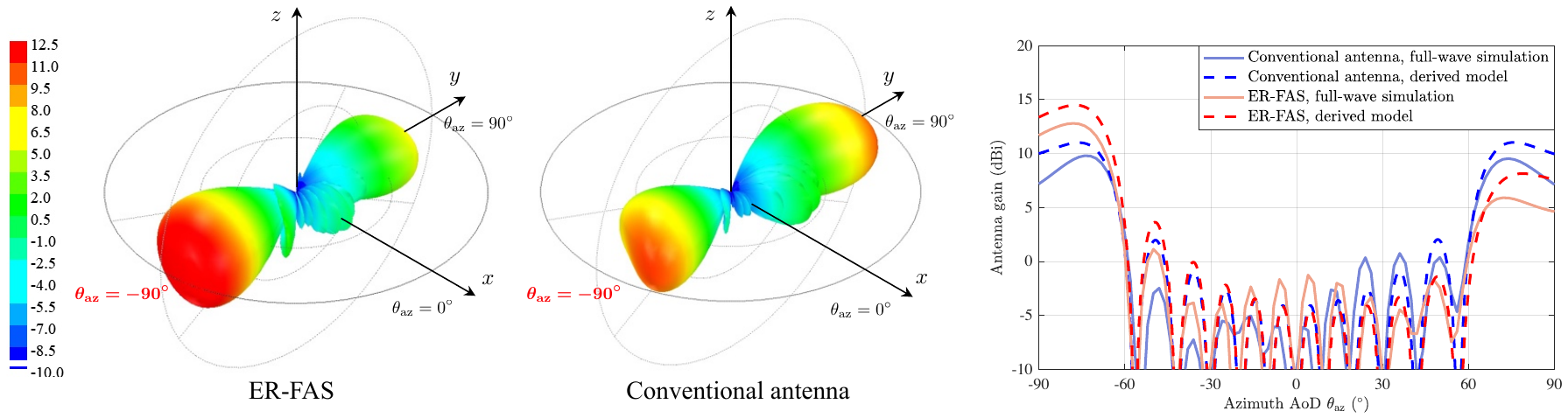}
  \vspace{-0.7em}
  \caption{ 
    The simulated array beampatterns (realized gain in dBi, i.e., relative to an isotropic radiator) for beamforming to $(\theta_\mathrm{az}=-90^\circ,\ \theta_\mathrm{el}=90^\circ)$ using the proposed beamforming algorithm. The conventional antenna stands for the case where each array element's radiation pattern is fixed. Here, we set this fixed radiation pattern as state 2 in Fig~\ref{fig_FAS_Element_Mechanism}. The same setup applies to Fig.~\ref{fig_RPs_60}.
   }
\label{fig_RPs_m90}
\vspace{-1em}
\end{figure*}

\begin{figure*}[t]
  \centering
  \includegraphics[width=0.9\linewidth]{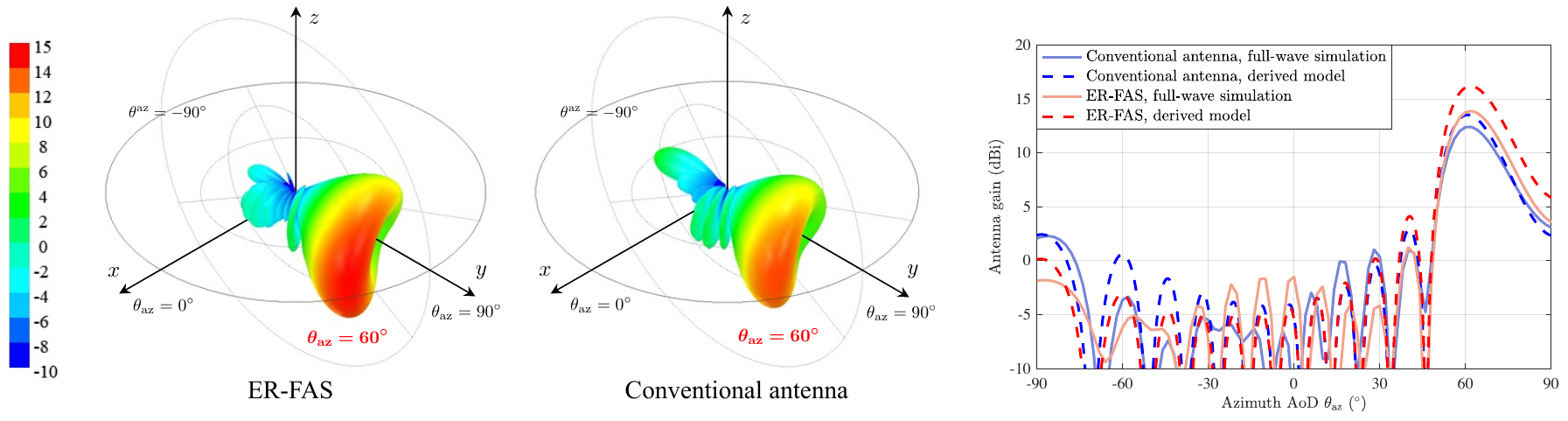}
  \vspace{-0.7em}
  \caption{ 
    The simulated array beampatterns for beamforming to  $(\theta_\mathrm{az}=60^\circ,\ \theta_\mathrm{el}=90^\circ)$.
   }
\label{fig_RPs_60}
\vspace{-1em}
\end{figure*}

This section presents the full-wave EM array beampattern simulation and the numerical evaluation of communication spectral efficiency. Throughout, the \ac{ER-FAS} selects radiation pattern of each element from three available states, namely, state 1, state 2, and state 3, shown in Fig.~\ref{fig_FAS_Element_Mechanism}. While more EM states are available, the simulation in this section will demonstrate that even with only three EM states for reconfigurability, substantial spectral efficiency enhancement can be achieved. Here, the conventional antennas used as the benchmark refer to array elements with fixed radiation patterns, specifically the typical broadside pattern defined as state 2 in Fig.~\ref{fig_FAS_Element_Mechanism}.

\subsection{Full-Wave EM Simulation Validation}\label{sec:Full-wave EM Simulation Validation}

To verify the accuracy of the communication model derived in Section~\ref{sec:channel modeling}, full-wave EM simulation of the designed ER-FAS in array configuration (as shown in Fig.~\ref{fig_FAS_Array}) is conducted using ANSYS HFSS. Considering the far-field improvement of the designed ER-FAS model, characterized by larger beam-scanning angles, two verification scenarios with large angles—specifically, the hemisphere border ($\theta_\mathrm{az}=-90^\circ$) and an arbitrary azimuthal angle ($\theta_\mathrm{az}=60^\circ$)—are selected to validate the design. It is worth noting that all angle information described for the designed ER-FAS is azimuthal, i.e., $\theta_\mathrm{el}=90^\circ$, considering the adopted 1D ER-FAS configuration. The phase shifts and radiation state of each array element are assigned using the beamforming algorithm~\ref{algo:1}. 

The simulated radiation patterns of the conventional antenna and the designed \ac{ER-FAS} for $\theta_\mathrm{az}=-90^\circ$ and $\theta_\mathrm{az}=60^\circ$ beamforming are shown in Fig.~\ref{fig_RPs_m90} and Fig.~\ref{fig_RPs_60}, respectively. Here, the unit of the ordinate is the antenna realized gain in decibels relative to an isotropic radiator (dBi), which is a standard metric for evaluating the far-field radiation characteristics of antennas. In the $XOY$ plane, the calculated results of the derived analytical model are also included for comparison with the full-wave simulation results. From the full-wave simulated 3D radiation patterns for $\theta_\mathrm{az}=-90^\circ$, it is evident that the overall radiated power of the developed ER-FAS design is more concentrated in the desired end-fire beam direction compared to the conventional antenna array based on both the numerical and full-wave EM simulation results. A similar observation can be made for $\theta_\mathrm{az}=60^\circ$ beamforming, where a higher gain is directed towards the desired beam direction. It is important to note that the phase states for both the conventional antenna and the designed element-reconfigurable array are maintained the same for a fair comparison.

To quantitatively demonstrate the gain enhancement of the proposed ER-FAS model, 2D $XOY$-plane radiation patterns are also presented, where the proposed model shows good agreement with the full-wave simulation. For the $-90^\circ$ beamforming direction, the proposed model demonstrates a gain enhancement of \unit[3.4]{dB}, which closely matches the \unit[4.5]{dB} enhancement observed in the full-wave simulation. Similarly, the gain enhancement at the $60^\circ$ beam direction is \unit[2.5]{dB} for the proposed analytical model, which is consistent with the full-wave simulated value of \unit[1.7]{dB}. The slight deviation between the analytical model and the full-wave simulation results is attributed to hardware impairments such as the mutual coupling effects~\cite{Zheng_pinjun2024MC}, which are outside the scope of this work but can be explored in future studies. Nevertheless, the proposed analytical model has been validated through full-wave simulation results, showing a good match. Additional far-field scenarios demonstrating the gain enhancement of the proposed ER-FAS compared to the conventional antenna are summarized in Table~\ref{Gain_Enhancement_Table}.

\begin{table}[t]
  \renewcommand{\arraystretch}{1.3}
  \begin{center}
  \caption{Gain Enhancement of the ER-FAS over Conventional Antenna}\vspace{-0.5em}
  \label{Gain_Enhancement_Table}
  {
  \begin{tabular}{ >{\centering\arraybackslash}m{1.5cm} !{\vrule width 1pt} 
                   >{\centering\arraybackslash}m{1cm} !{\color{black}\vrule} 
                   >{\centering\arraybackslash}m{1cm} !{\color{black}\vrule} 
                   >{\centering\arraybackslash}m{1cm} !{\color{black}\vrule} 
                   >{\centering\arraybackslash}m{1cm} !{\color{black}\vrule} 
                   >{\centering\arraybackslash}m{1cm} }
    \Xhline{1pt}
    \multirow{2}{*}{\textbf{Gain (dBi)}} & \multicolumn{5}{c}{\textbf{Beamforming direction}} \\
    \cline{2-6}
     & $\bm{\pm 50^\circ}$ & $\bm{\pm 60^\circ}$ & $\bm{\pm 70^\circ}$ & $\bm{\pm 80^\circ}$ & $\bm{\pm 90^\circ}$ \\
    \Xhline{1pt}
    Conventional antenna & 13.6 & 12.5 & 11.1 & 9.2 & 7.1 \\
    \hline
    ER-FAS               & 14.7 & 14.0 & 13.4 & 12.7 & 11.6 \\
    \hline
    \textbf{Gain enhancement} & \textbf{1.1} & \textbf{1.5} & \textbf{2.3} & \textbf{3.5} & \textbf{4.5} \\
    \Xhline{1pt}
  \end{tabular}
  }
  \end{center}
\end{table}

\subsection{Numerical Spectral Efficiency Evaluation}\label{sec:Beamforming Numerical Evaluation}

\begin{figure}[t]
  \centering
  \includegraphics[width=0.85\linewidth]{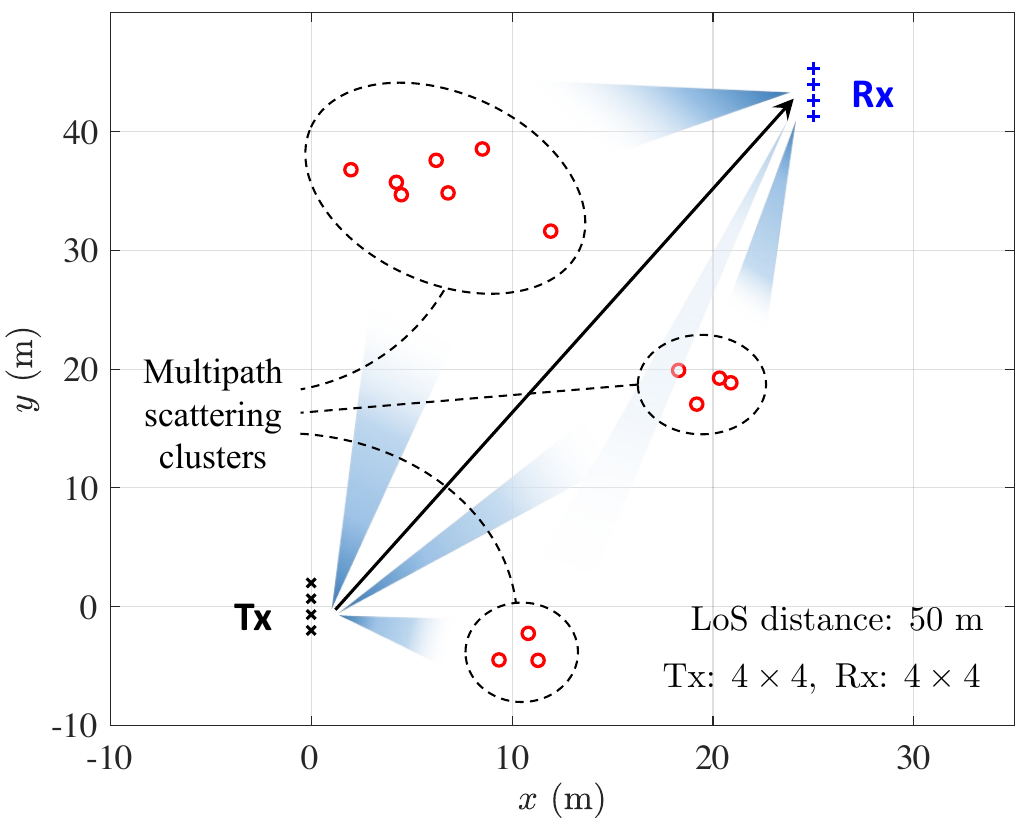}
  \vspace{-0.5em}
  \caption{ 
     Illustration of the adopted \ac{MIMO} communication setup based on the \ac{SV} model, featuring a \ac{LoS} path and multiple \ac{NLoS} paths propagating through 3 clusters of scatterers.
    }
  \label{fig_simSetup}
\end{figure}

After validating the proposed ER-FAS model through full-wave EM simulation, the assessment of the communication spectral efficiency is presented in this subsection. In this subsection, we first evaluate the communication performance of the proposed system and beamforming algorithm in far-field scenarios. The simulation setup is illustrated in Fig.~\ref{fig_simSetup}. Specifically, we consider a far-field communication system with a $4\times 4$ transmitter and a $4\times 4$ receiver, both equipped with \ac{ER-FAS}. The physical wireless channels are generated according to the \ac{SV} model, with 3 scatterer clusters generating \ac{NLoS} multipath. The signal frequency is configured to \unit[$3.55$]{GHz}, and the thermal noise at the receiver is set to \unit[$-95$]{dBm} based on experimental measurements in~\cite[Fig.~25]{Wang_ruiqi2024TAP}. 

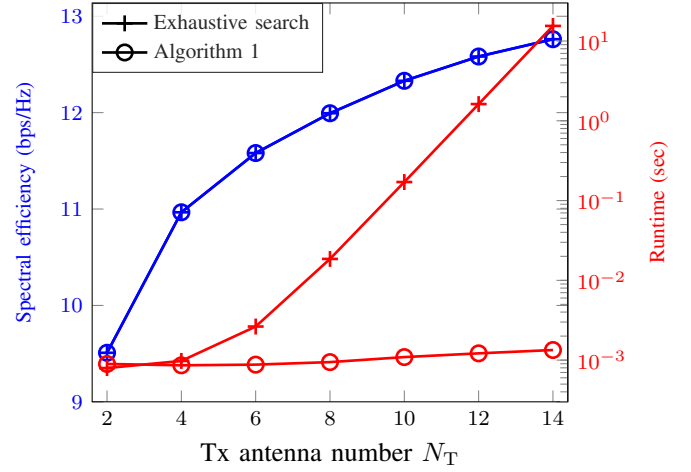
\begin{figure}[t]
    \centering
    % This file was created by matlab2tikz.
%
%The latest updates can be retrieved from
%  http://www.mathworks.com/matlabcentral/fileexchange/22022-matlab2tikz-matlab2tikz
%where you can also make suggestions and rate matlab2tikz.
%
\begin{tikzpicture}

\begin{axis}[
    width=3.1in,
    height=2.7in,
    at={(0in,0in)},
    xlabel={Tx antenna number $N_\mathrm{T}$},
    xtick=data,
    xlabel style={font=\color{white!15!black},font=\footnotesize},
    xticklabel style = {font=\color{white!15!black},font=\footnotesize},
    axis y line*=left,
    axis x line*=bottom,
    axis lines=box,
    ymin=9,
    xmin = 1.6,
    xmax = 14.4,
    ylabel style={color=blue,font=\footnotesize},
    yticklabel style = {color=blue,font=\footnotesize},
    tick align=inside,
    ylabel style={color=black, at={(axis description cs:-0.10,0.5)}, anchor=south},
    ylabel={\blue{Spectral efficiency (bps/Hz)}},
    legend style={at={(0.03,0.97)}, anchor=north west}
]

\addplot+[blue, mark=o, thick, line width=1pt, ,mark size=2.9pt] coordinates {(2,9.5086) (4,10.9662) (6,11.5809) (8,11.9938) (10,12.3303) (12,12.5824) (14,12.7618)};
%\addlegendentry{Algo 1};

\addplot+[blue, mark=+, thick, line width=1pt, ,mark size=3pt] coordinates {(2,9.5086) (4,10.9662) (6,11.5809) (8,11.9938) (10,12.3303) (12,12.5824) (14,12.7618)};
%\addlegendentry{Exhaustive search};

\end{axis}

\begin{axis}[
    width=3.1in,
    height=2.7in,
    at={(0in,0in)},
    axis y line*=right,
    axis x line=none,
    ymode=log,
    ymin=3e-4, ymax=30,
    xmin = 1.6,
    xmax = 14.4,
    yticklabel style={color=red},
    tick align=inside,
    ylabel style={color=red, at={(axis description cs:1.23,0.5)}, anchor=south, font=\footnotesize},
    yticklabel style = {color=red,font=\footnotesize},
    ylabel={\red{Runtime (sec)}},
    legend style={at={(0,1)}, anchor=north west, font=\footnotesize, legend cell align=left, align=left, draw=white!15!black, fill opacity=0.85}
]

\addplot+[black, mark=+, thick, line width=1pt, ,mark size=3pt] coordinates {(-1,0.000795155555555556) (0,0.000976663888888889)};
\addlegendentry{Exhaustive search};

\addplot+[black, mark=o, thick, line width=1pt, ,mark size=2.9pt] coordinates {(-1,0.000795155555555556) (0,0.000976663888888889)};
\addlegendentry{Algorithm 1};

\addplot+[red, mark=o, thick, line width=1pt, ,mark size=2.9pt] coordinates {(2,0.000895043055555556) (4,0.000860661111111111) (6,0.000877512500000000) (8,0.000943893055555556) (10,0.00109066805555556) (12,0.00121528611111111) (14,0.00133832222222222)};
%\addlegendentry{Algo 1 runtime};

\addplot+[red, mark=+, thick, line width=1pt, ,mark size=3pt] coordinates {(2,0.000795155555555556) (4,0.000976663888888889) (6,0.00263505277777778) (8,0.0185501291666667) (10,0.170748719444444) (12,1.61969631250000) (14,15.4621011958333)};
%\addlegendentry{Search runtime};

\end{axis}
\end{tikzpicture}
    \vspace{-3em}
    \caption{Comparison between the proposed Algorithm~\ref{algo:1} and the exhaustive search solution in terms of system spectral efficiency and algorithm runtime.}
    \label{fig_search}
\end{figure}

In Fig.~\ref{fig_search}, we first compare the proposed Algorithm~\ref{algo:1} with the exhaustive search solution in terms of achievable spectral efficiency and algorithm runtime across different numbers of transmit antennas $N_\mathrm{T}$. The results are averaged from 100 Monte Carlo simulations using MATLAB R2024a on a MacBook Pro 18.3 (with Apple M1 Pro chip). The results show that Algorithm~\ref{algo:1} consistently identifies the global optimum in all tested cases, achieving the same spectral efficiency as the exhaustive search method. However, its runtime is significantly lower, especially for larger antenna arrays (e.g., when $N_\mathrm{T} \geq 6$). This aligns with the complexity analysis in Sec.~\ref{sec:complexity}. It should be noted, however, that this ideal performance is facilitated by the relatively simple single-user communication setup considered in this paper. In practical \ac{MIMO} systems, which often serve multiple users simultaneously and are subject to mutual interference, achieving the global optimum becomes more challenging and requires more advanced solutions.

\begin{figure}[t]
    \centering
    % This file was created by matlab2tikz.
%
%The latest updates can be retrieved from
%  http://www.mathworks.com/matlabcentral/fileexchange/22022-matlab2tikz-matlab2tikz
%where you can also make suggestions and rate matlab2tikz.
%
\begin{tikzpicture}[scale=1]

\begin{axis}[%
width=2.8in,
height=1.6in,
at={(0in,0in)},
scale only axis,
xmin=-30.8,
xmax=20.8,
xlabel style={font=\color{white!15!black},font=\footnotesize},
xticklabel style = {font=\color{white!15!black},font=\footnotesize},
xlabel={Transmit power (dBm)},
ymin=-0.5,
ymax=21,
ylabel style={font=\color{white!15!black},font=\footnotesize},
yticklabel style = {font=\color{white!15!black},font=\footnotesize},
ylabel={Spectral efficiency (bps/Hz)},
axis background/.style={fill=white},
xmajorgrids,
ymajorgrids,
legend style={at={(0,1)}, anchor=north west, font=\scriptsize, legend cell align=left, align=left, draw=white!15!black, fill opacity=0.85}
]
\addplot [color=red, line width=1.2pt, mark=o, mark options={solid, red}]
  table[row sep=crcr]{%
-30	3.68849231989793\\
-25	5.27083899339729\\
-20	6.90602322372848\\
-15	8.55873816521936\\
-10	10.217083769551\\
-5	11.8772188025636\\
0	13.5379205934543\\
5	15.1988016981716\\
10	16.8597395158298\\
15	18.5206952685878\\
20	20.1816566930116\\
};
\addlegendentry{\ac{ER-FAS}: optimized states + phase shifts}

\addplot [color=blue, line width=1.2pt, mark=diamond, mark options={solid, blue}, mark size=2.5pt]
  table[row sep=crcr]{%
-30	2.10394787607138\\
-25	3.51498728691089\\
-20	5.08697112662336\\
-15	6.7186153947728\\
-10	8.37018231995546\\
-5	10.0281619526709\\
0	11.6881809590367\\
5	13.3488460293724\\
10	15.0097155190572\\
15	16.6706496634222\\
20	18.3316042545532\\
};
\addlegendentry{Conventional antenna: optimized phase shifts}

\addplot [color=black, line width=1.2pt, mark=square, mark options={solid, black}, mark size=1.8pt]
  table[row sep=crcr]{%
-30	0.0215937201346879\\
-25	0.067208853949102\\
-20	0.202639659929333\\
-15	0.562550520380571\\
-10	1.32654897420942\\
-5	2.52826590133563\\
0	4.00721363456122\\
5	5.605487999915\\
10	7.24604718876632\\
15	8.90049812642944\\
20	10.5593964133262\\
};
\addlegendentry{Conventional antenna: random phase shifts}

\end{axis}

\end{tikzpicture}%
    \vspace{-3em}
    \caption{Far-field spectral efficiency versus transmit power.}
    \label{fig_SE_vs_PT}
\end{figure}
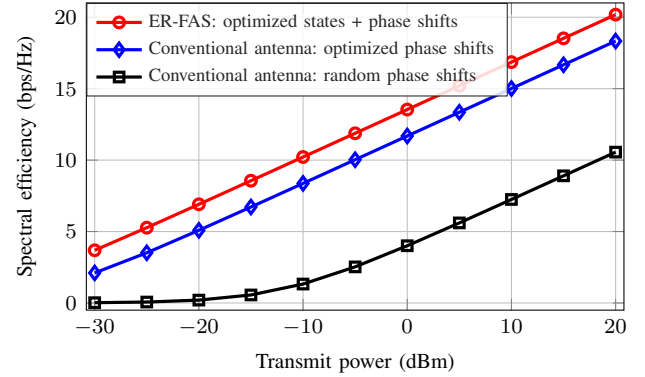

Fig.~\ref{fig_SE_vs_PT} compares the communication spectral efficiency of the \ac{ER-FAS} and the conventional antenna system. The conventional antenna system serves as a benchmark, adopting the same array configuration as the \ac{ER-FAS}, but with each antenna having a static radiation pattern, which is chosen as State~2 in Fig.~\ref{fig_FAS_Element_Mechanism}. In the far-field, its channel model is described by the conventional \ac{SV} model~$\Hm_\mathrm{CV}$ in~\eqref{eq:HFF_conv}, which can also be recovered from $\Hm_\mathrm{ER}$ by setting all antenna gains to unity, i.e., $G_{\Tt,i}(\thetav) = G_{\Rt,j}(\phiv) = 1$, $\forall i,j$. The conventional near-field channel model used in the subsequent evaluations is obtained in the same way. The spectral efficiency is calculated using~\eqref{eq:Rate}. The results in Fig.~\ref{fig_SE_vs_PT} show that both the \ac{ER-FAS} and the conventional antenna system require beamforming optimization to achieve high spectral efficiency, as random phase shifts result in significantly lower performance. Furthermore, the \ac{ER-FAS} provides an approximate \unit[$1.5$]{bps/Hz} gain in spectral efficiency compared to the conventional antenna, demonstrating its strong potential for enhancing communication performance.

\begin{figure}[t]
    \centering
    % This file was created by matlab2tikz.
%
%The latest updates can be retrieved from
%  http://www.mathworks.com/matlabcentral/fileexchange/22022-matlab2tikz-matlab2tikz
%where you can also make suggestions and rate matlab2tikz.
%
\begin{tikzpicture}[scale=1]

\begin{axis}[%
width=2.8in,
height=1.6in,
at={(0in,0in)},
scale only axis,
xmin=-92,
xmax=92,
xtick = {-90,-60,-30,0,30,60,90},
xlabel style={font=\color{white!15!black},font=\footnotesize},
xticklabel style = {font=\color{white!15!black},font=\footnotesize},
xlabel={Azimuth AoD $\theta_\mathrm{az}$ ($^\circ$)},
ymin=12.8,
ymax=18.2,
ylabel style={font=\color{white!15!black},font=\footnotesize},
yticklabel style = {font=\color{white!15!black},font=\footnotesize},
ylabel={Spectral efficiency (bps/Hz)},
axis background/.style={fill=white},
xmajorgrids,
ymajorgrids,
legend style={at={(0.11,0.25)}, anchor=north west, font=\scriptsize, legend cell align=left, align=left, draw=white!15!black, fill opacity=0.85}
]
\addplot [color=red, line width=1.2pt, mark=o, mark options={solid, red}]
  table[row sep=crcr]{%
-90	15.3034904630681\\
-85	15.7065578782098\\
-80	16.0746745859682\\
-75	16.4036873695591\\
-70	16.6929910636375\\
-65	16.9444455546269\\
-60	17.1613193556522\\
-55	17.3474457954974\\
-50	17.5065851792742\\
-45	17.6419331237727\\
-40	17.7558680779325\\
-35	17.8497852182234\\
-30	17.9240432924517\\
-25	17.9780044794582\\
-20	18.0100676066093\\
-15	18.0177213000231\\
-10	17.9976237113095\\
-5	17.9744566740116\\
0	17.9882359785008\\
5	17.9768351652998\\
10	17.9580397694751\\
15	17.9858375556482\\
20	17.9871331026165\\
25	17.9649626405019\\
30	17.92128610323\\
35	17.8569871279211\\
40	17.7719594253262\\
45	17.6651333570606\\
50	17.5346221032831\\
55	17.3777947504843\\
60	17.1914756152978\\
65	16.9722033678006\\
70	16.7166296826604\\
75	16.4220839964978\\
80	16.087390743277\\
85	15.7137995463701\\
90	15.3061080777354\\
};
\addlegendentry{\ac{ER-FAS}: optimized states + phase shifts}

\addplot [color=blue, line width=1.2pt, mark=diamond, mark options={solid, blue},mark size=2.5pt]
  table[row sep=crcr]{%
-90	13.0310097832611\\
-85	13.3236101064058\\
-80	13.6788223955195\\
-75	14.0938206439362\\
-70	14.5510184500146\\
-65	15.025479238905\\
-60	15.4927230333596\\
-55	15.9334682159674\\
-50	16.334924270357\\
-45	16.6901209448807\\
-40	16.9965196939107\\
-35	17.2546649184519\\
-30	17.4670478063509\\
-25	17.6372625157446\\
-20	17.7693550453466\\
-15	17.8673382124196\\
-10	17.9347664333847\\
-5	17.9744566740116\\
0	17.9882359785008\\
5	17.9768351652998\\
10	17.939775880993\\
15	17.8754436831681\\
20	17.7811877004739\\
25	17.6535066648946\\
30	17.4883545364138\\
35	17.281479352143\\
40	17.0289547545375\\
45	16.7277912578202\\
50	16.3768132841588\\
55	15.9777953269248\\
60	15.5368173650064\\
65	15.065779678896\\
70	14.5832531122255\\
75	14.1136308333984\\
80	13.6829339650131\\
85	13.3111952842752\\
90	13.0044574921508\\
};
\addlegendentry{Conventional antenna: optimized phase shifts}

\end{axis}
\end{tikzpicture}%
    \vspace{-3em}
    \caption{Far-field spectral efficiency versus azimuth \ac{AoD} $\theta_\mathrm{az}$, with the elevation \ac{AoD} fixed at $\theta_\mathrm{el}=\unit[90]{^\circ}$.}
    \label{fig_SE_vs_AoD}
\end{figure}
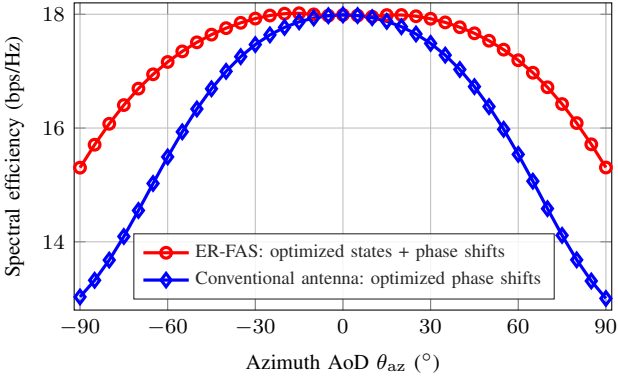

To comprehensively evaluate the performance of the proposed system, Fig.~\ref{fig_SE_vs_AoD} further presents the communication spectral efficiency as a function of the \ac{AoD}. In this trial, we keep the Tx-Rx \ac{LoS} distance fixed while varying the Rx locations to correspond to different azimuth \acp{AoD}, while the elevation \ac{AoD} is fixed to $\unit[90]{^\circ}$. To better illustrate the underlying trend, we consider only the \ac{LoS} path, eliminating the impact of \ac{NLoS} multipath. Note that the effect of multipath will be analyzed in a later trial. As shown in Fig.~\ref{fig_SE_vs_AoD}, the proposed \ac{ER-FAS} achieves spectral efficiency comparable to that of the conventional antenna system at small azimuth \acp{AoD}. However, at larger azimuth \acp{AoD}, it significantly outperforms the conventional antenna system, aligning well with the full-wave simulation results.

\begin{figure}[t]
    \centering
    % This file was created by matlab2tikz.
%
%The latest updates can be retrieved from
%  http://www.mathworks.com/matlabcentral/fileexchange/22022-matlab2tikz-matlab2tikz
%where you can also make suggestions and rate matlab2tikz.
%
\begin{tikzpicture}[scale=1]

\begin{axis}[%
width=2.8in,
height=1.6in,
at={(0in,0in)},
scale only axis,
xmin=-0.2,
xmax=18.2,
xtick = {0,3,6,9,12,15,18},
xlabel style={font=\color{white!15!black},font=\footnotesize},
xticklabel style = {font=\color{white!15!black},font=\footnotesize},
xlabel={Number of multipaths per cluster},
ymin=11,
ymax=18,
ylabel style={font=\color{white!15!black},font=\footnotesize},
yticklabel style = {font=\color{white!15!black},font=\footnotesize},
ylabel={Spectral efficiency (bps/Hz)},
axis background/.style={fill=white},
xmajorgrids,
ymajorgrids,
legend style={at={(1,1)}, anchor=north east, font=\scriptsize, legend cell align=left, align=left, draw=white!15!black, fill opacity=0.85}
]
\addplot [color=red, line width=1.2pt, mark=o, mark options={solid, red}]
  table[row sep=crcr]{%
0	16.087390743277\\
3	12.9669627278226\\
6	12.4495852281104\\
9	12.2753091367489\\
12	12.2059641361206\\
15	12.1864293422317\\
18	12.1569004774112\\
};
\addlegendentry{\ac{ER-FAS}: optimized states + phase shifts}

\addplot [color=blue, line width=1.2pt, mark=diamond, mark options={solid, blue},mark size=2.5pt]
  table[row sep=crcr]{%
0	13.6829339650131\\
3	12.1195872950809\\
6	11.9614666732888\\
9	11.9791287537761\\
12	11.9252389244574\\
15	11.9448406312366\\
18	11.9098829011175\\
};
\addlegendentry{Conventional antenna: optimized phase shifts}

\addplot [color=red, dashed, line width=1.2pt, mark=o, mark options={solid, red}, forget plot]
  table[row sep=crcr]{%
0	17.7719594253263\\
3	14.6172110140971\\
6	13.8397487542046\\
9	13.7105935082057\\
12	13.3896920399889\\
15	13.2605223249006\\
18	13.1098718325458\\
};
\addplot [color=blue, dashed, line width=1.2pt, mark=diamond, mark options={solid, blue},mark size=2.5pt, forget plot]
  table[row sep=crcr]{%
0	17.0289547545375\\
3	13.9329371174651\\
6	13.1823217500994\\
9	13.0402743464238\\
12	12.7512325313032\\
15	12.6298635913446\\
18	12.4830501987156\\
};
\end{axis}

\begin{axis}[%
width=2.8in,
height=1.6in,
at={(0in,0in)},
scale only axis,
xmin=-0.2,
xmax=18.2,
ymin=11,
ymax=18,
axis line style={draw=none},
ticks=none,
axis x line*=bottom,
axis y line*=left
  ]
  \draw [black] (axis cs:7.5,13.5) ellipse [x radius=0.6, y radius=0.7];
  \node[right, align=left]
    at (axis cs:7.5,14.5) {\footnotesize{$\theta_\mathrm{az}=40^\circ$}};
  \draw [black] (axis cs:10.5,12.1) ellipse [x radius=0.55, y radius=0.5];
  \node[right, align=left]
    at (axis cs:10.5,11.4) {\footnotesize{$\theta_\mathrm{az}=80^\circ$}};
  \end{axis}
  
\end{tikzpicture}%
    \vspace{-3em}
    \caption{Far-field spectral efficiency versus the number of \ac{NLoS} multipath.}
    \label{fig_SE_vs_numNLoS}
\end{figure}
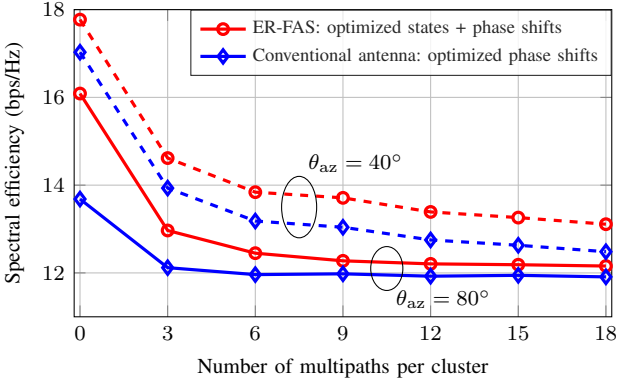

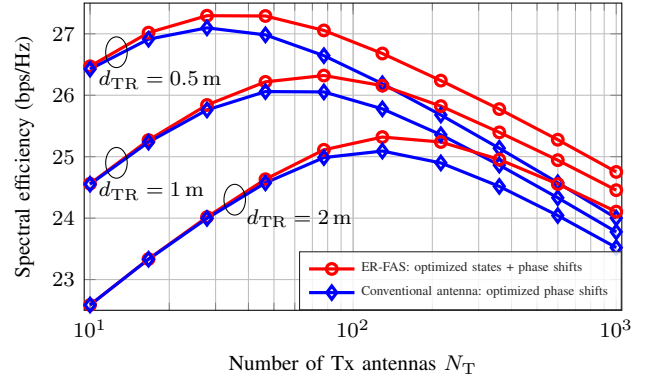
\begin{figure}[t]
    \centering
    % This file was created by matlab2tikz.
%
%The latest updates can be retrieved from
%  http://www.mathworks.com/matlabcentral/fileexchange/22022-matlab2tikz-matlab2tikz
%where you can also make suggestions and rate matlab2tikz.
%
\begin{tikzpicture}[scale=1]

\begin{axis}[%
width=2.8in,
height=1.6in,
at={(0in,0in)},
scale only axis,
xmode=log,
xmin=9.5,
xmax=1050,
xminorticks=true,
xlabel style={font=\color{white!15!black},font=\footnotesize},
xticklabel style = {font=\color{white!15!black},font=\footnotesize},
xlabel={Number of Tx antennas $N_\mathrm{T}$},
ymin=22.5,
ymax=27.5,
ytick={23,24,25,26,27},
ylabel style={font=\color{white!15!black},font=\footnotesize},
yticklabel style = {font=\color{white!15!black},font=\footnotesize},
ylabel={Spectral efficiency (bps/Hz)},
axis background/.style={fill=white},
xmajorgrids,
xminorgrids,
ymajorgrids,
legend style={at={(1,0)}, anchor=south east, font=\tiny, legend cell align=left, align=left, draw=white!15!black, fill opacity=0.85}
]
\addplot [color=red, line width=1.2pt, mark=o, mark options={solid, red}]
  table[row sep=crcr]{%
10	26.4712074628579\\
16.6810053720006	27.0188465902714\\
27.8255940220712	27.2946548342627\\
46.4158883361278	27.2902027353343\\
77.4263682681127	27.0550905206984\\
129.154966501488	26.6795157626272\\
215.443469003188	26.2386358575941\\
359.381366380463	25.7745863196871\\
599.484250318941	25.2759800666057\\
1000	24.7519238723694\\
};
\addlegendentry{\ac{ER-FAS}: optimized states + phase shifts}

\addplot [color=blue, line width=1.2pt, mark=diamond, mark options={solid, blue},mark size=2.5pt]
  table[row sep=crcr]{%
10	26.4251410150594\\
16.6810053720006	26.9124107721767\\
27.8255940220712	27.0970919775509\\
46.4158883361278	26.9828989324395\\
77.4263682681127	26.64429843101\\
129.154966501488	26.1875711697996\\
215.443469003188	25.6789505164911\\
359.381366380463	25.1371679246426\\
599.484250318941	24.5762763069165\\
1000	24.0003295892389\\
};
\addlegendentry{Conventional antenna: optimized phase shifts}

\addplot [color=red, line width=1.2pt, mark=o, mark options={solid, red}, forget plot]
  table[row sep=crcr]{%
10	24.560080062499\\
16.6810053720006	25.2701020134103\\
27.8255940220712	25.841907369234\\
46.4158883361278	26.2182939609466\\
77.4263682681127	26.3212892938695\\
129.154966501488	26.1563148459005\\
215.443469003188	25.8234038481682\\
359.381366380463	25.3961151571643\\
599.484250318941	24.9432685699741\\
1000	24.454865847056\\
};
\addplot [color=blue, line width=1.2pt, mark=diamond, mark options={solid, blue},mark size=2.5pt, forget plot]
  table[row sep=crcr]{%
10	24.5525383673367\\
16.6810053720006	25.2372994123218\\
27.8255940220712	25.7610042831001\\
46.4158883361278	26.0599789064853\\
77.4263682681127	26.0538741709972\\
129.154966501488	25.7785896360876\\
215.443469003188	25.3572855787501\\
359.381366380463	24.8628301423659\\
599.484250318941	24.3311554317274\\
1000	23.7752113970412\\
};
\addplot [color=red, line width=1.2pt, mark=o, mark options={solid, red}, forget plot]
  table[row sep=crcr]{%
10	22.586611857715\\
16.6810053720006	23.3339937603806\\
27.8255940220712	24.0170144336728\\
46.4158883361278	24.6329027115887\\
77.4263682681127	25.1121095535015\\
129.154966501488	25.3198681829862\\
215.443469003188	25.2389899418525\\
359.381366380463	24.954298477285\\
599.484250318941	24.5546029105114\\
1000	24.1060924608517\\
};
\addplot [color=blue, line width=1.2pt, mark=diamond, mark options={solid, blue},mark size=2.5pt, forget plot]
  table[row sep=crcr]{%
10	22.586611857715\\
16.6810053720006	23.3308623927535\\
27.8255940220712	23.9957263179188\\
46.4158883361278	24.5738395506945\\
77.4263682681127	24.985570798367\\
129.154966501488	25.0917645631999\\
215.443469003188	24.8982024896258\\
359.381366380463	24.5165057649575\\
599.484250318941	24.0427476503587\\
1000	23.520955600272\\
};
\end{axis}

\begin{axis}[%
width=2.8in,
height=1.6in,
at={(0in,0in)},
scale only axis,
xmin=0,
xmax=100,
ymin=0,
ymax=100,
axis line style={draw=none},
ticks=none,
axis x line*=bottom,
axis y line*=left
  ]
  \draw [black] (axis cs:6,84) ellipse [x radius=2, y radius=5];
  \node[right, align=left]
    at (axis cs:1,75) {\footnotesize{$d_\mathrm{TR}=\unit[0.5]{m}$}};
  \draw [black] (axis cs:6,48) ellipse [x radius=2, y radius=5];
  \node[right, align=left]
    at (axis cs:1,39) {\footnotesize{$d_\mathrm{TR}=\unit[1]{m}$}};
  \draw [black] (axis cs:28,36) ellipse [x radius=2, y radius=5];
  \node[right, align=left]
    at (axis cs:28.5,30) {\footnotesize{$d_\mathrm{TR}=\unit[2]{m}$}};
  \end{axis}

\end{tikzpicture}%
    \vspace{-3em}
    \caption{Near-field spectral efficiency versus Tx array size. The Tx is set as an $N_\mathrm{T}\times 1$ linear array along the $Y$-axis, with the Rx located in front of the transmitter at varying Tx-Rx distances $d_\mathrm{TR}$. The Tx array spacing is fixed as half-wavelength.}
    \label{fig_SE_NF}
\end{figure}

Now, we assess the impact of \ac{NLoS} multipath on system performance. We evaluate spectral efficiency at two azimuth \acp{AoD}, $\theta_\mathrm{az}=\unit[40]{^\circ}$ and $\theta_\mathrm{az}=\unit[80]{^\circ}$, while keeping the elevation \ac{AoD} fixed at $\unit[90]{^\circ}$. For each \ac{AoD}, we examine spectral efficiency under various numbers of multipath components by adjusting the number of scatterers per cluster. In both \ac{AoD} cases, we observe that an increasing number of \ac{NLoS} paths leads to a decline in spectral efficiency. Additionally, the \ac{ER-FAS} consistently outperforms the conventional antenna system. However, in the large \ac{AoD} case, the performance advantage of \ac{ER-FAS} over the conventional benchmark diminishes in the presence of rich \ac{NLoS} multipath. In contrast, for small \ac{AoD}, \ac{ER-FAS} maintains a steady performance gain over the conventional antenna system. This observation implies that the significant performance gain of \ac{ER-FAS} at large \ac{AoD} is particularly advantageous in sparse multipath environments, such as mmWave and THz communication systems, while its benefits may be less pronounced in rich-scattering conditions.

Furthermore, we evaluate the system performance in near-field scenarios. In this simulation, the transmitter is configured as an $N_\mathrm{T}\times 1$ linear array along the $Y$-axis. The \ac{AoD} is fixed at $\theta_\mathrm{az}=\unit[0]{^\circ}$ and $\theta_\mathrm{el}=\unit[90]{^\circ}$, i.e., the Rx is located in front of the Tx. Various Tx-Rx distances $d_\mathrm{TR}$ are tested, with the Tx array spacing fixed at half-wavelength. As shown in Fig.~\ref{fig_SE_NF}, we observe a performance gain of \ac{ER-FAS} compared to the conventional benchmark as the Tx array size increases and the Tx-Rx distance decreases. This indicates that the proposed \ac{ER-FAS} is particularly beneficial in near-field scenarios, especially when large antenna arrays are used. The performance improvement stems from the significant degree of freedom enabled by the radiation pattern reconfigurability, which allows for more effective focusing of radiated power in the desired focal regions. Furthermore, note that in these evaluations, the total transmit power is kept constant. This can be verified from the constraint in~\eqref{eq:Ddef}, where we always have $\|\fv\|_2 = \|\wv\|_2 = 1$ across different antenna array sizes. Although in practice each antenna element is typically equipped with its own power amplifier, which implies that more antennas always result in higher power consumption, our theoretical analysis adopts a constant-power assumption to isolate the effects of beamforming and antenna reconfigurability. Therefore, the result shows a similar trend to the energy efficiency in massive \ac{MIMO} systems~\cite[Fig.~2]{Prasad2017Energy}. The system spectral efficiency begins to degrade with increasing $N_\mathrm{T}$ when the array is excessively large, suggesting that oversized arrays may lead to energy waste. In our simulation scenario, when the array operates in the near-field and becomes overly large, antenna elements near the array edges experience longer propagation distances and, therefore, greater path loss than those at the center. This results in a reduction in overall spectral efficiency under a fixed total transmit power. Nevertheless, the radiation pattern reconfigurability enabled by \ac{ER-FAS} can partially compensate for this degradation.

\section{Hardware Fabrication}\label{sec:Hardware}

In this section, hardware prototypes are fabricated for both the ER-FAS array element and the entire array. The element prototype validates the software-controlled fluidic manipulation capability and radiation pattern reconfigurability using liquid metal, while the array prototype verifies the beamforming radiation pattern and gain enhancement in comparison to full-wave simulations

\subsection{ER-FAS Array Element}

\begin{figure}[t]
  \centering 
  \includegraphics[width=\columnwidth]{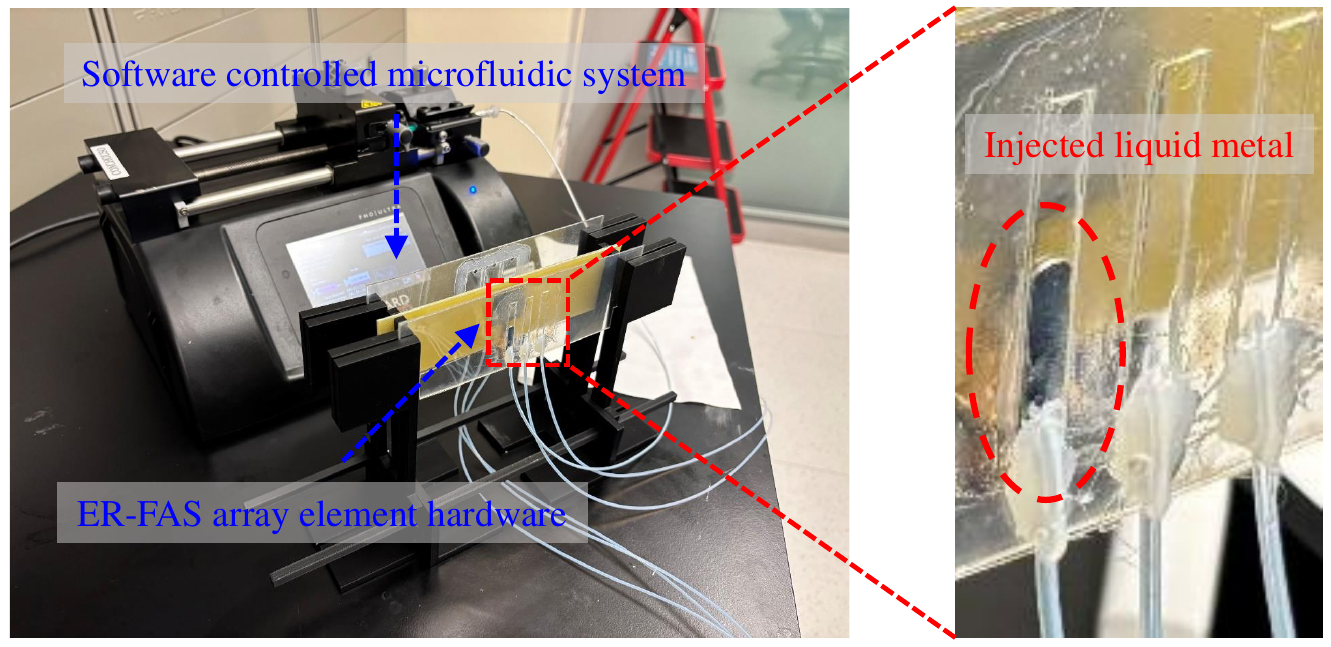}
  \vspace{-1.8em}
  \caption{
    The software-controllable ER-FAS array element prototype.
  }
  \label{fig_Array_Element_Prototype}
\end{figure}

Firstly, a software-controllable ER-FAS array element utilizing liquid metal is fabricated to validate both the radiation pattern reconfigurability and the dynamic fluidic manipulation capability, which is shown in Fig.~\ref{fig_Array_Element_Prototype}. In the array element fabrication, the Galinstan liquid metal alloy, consisting of 68.5\% gallium (Ga), 21.5\% indium (In), and 10\% tin (Sn), is adopted to demonstrate the antenna reconfigurability enabled by the liquid metal. It exhibits a low viscosity of 2.4 mPa$\cdot$s (20$^\circ$C), which is close to water’s viscosity of around 1 mPa$\cdot$s, demonstrating its decent fluidity. Meanwhile, it has a low melting point of -19°C with a conductivity of 3.46 × $10^\text{6}$ S/m, ensuring its fluid characteristic at regular room temperature with satisfactory electrical conductivity properties. The electrical conductivity is sufficient to meet the RF requirements of the ER-FAS. Full-wave simulations show a conductor loss of 1.38 mW under 1 W input, indicating minimal impact on antenna efficiency.

\begin{figure*}[t]
  \centering
  \includegraphics[width=0.85\linewidth]{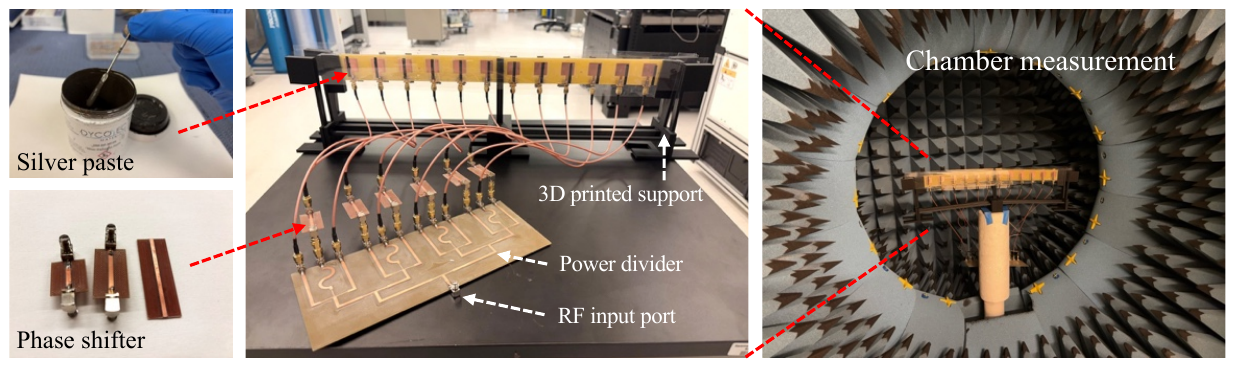}
  \vspace{-0.7em}
  \caption{ 
     Fabricated ER-FAS array prototype and the measurement setup in an anechoic chamber.
    }
  \label{fig_fabricated_FAS_Array}
  \vspace{-1em}
\end{figure*}

Then, the independent channels as containers of the liquid metal are fabricated. Specifically, the top layer is a 125 $\mu$m PMMA sheet with inlet and outlet cuts, the middle layer is a 0.4 mm PET substrate with metallic patterns, and the bottom layer is a full 125 $\mu$m PMMA sheet. All these different layers are cut through a laser cutting machine VLS 3.50. Then, these three layers are sandwiched together through adhesives.

Finally, the excitation source of the planar monopole antenna is fabricated with two single-clad 1.6 mm-thick FR4 substrates etched through an LPKF laser electronics equipment. Two FR4 substrates are assembled using fine screws with a diameter of 0.8 mm to avoid potential influence on antenna array performance. The antenna port is soldered with an SMA connector that is used to receive the RF signal for the entire array element. After all the prototypes have been prepared, the overall array element designs are assembled and fixed through customized 3D printed supports. During this process, PLA filament is adopted using a Raise3D Pro2 printer. Finally, a software-controllable pumping machine of the Harvard Apparatus type is deployed to control the injection or extraction of the Galinstan alloy.  According to the fabricated channels, a pumping speed of 50 $\mu$l/min is determined and set in the machine software for liquid injection through a specified syringe. Since the fluid control components of the designed ER-FAS prototype are parasitic reflectors and directors that are not directly connected to the planar monopole source antenna, the system exhibits robust RF performance and is not significantly affected by mechanical disturbances such as vibration or temperature variations. While this proof-of-concept uses a standard lab pump, future systems are expected to adopt compact, low-power piezoelectric micropumps (e.g., 30 × 15 mm, 50 mW~\cite{bartels2025micropumps}). Their small size supports scalable multi-antenna integration. With flow rates up to 14 mL/min, they enable millisecond-level reconfigurability suitable for indoor channel coherence times. For practical use, inert gas operation and sealed channels are recommended to prevent liquid metal oxidation.

\subsection{ER-FAS Array}

To further validate the designed ER-FAS concept and model, an ER-FAS array configuration has been fabricated. The fabrication details are provided as follows. First, the planar monopole antenna is similar to the array element manufacturing, but with a larger number of elements of 6 for the PCB sheet size. The designed ER-FAS model consists of two \( 1 \times 6 \) sub-arrays, and the final array design follows a \( 1 \times 12 \) array configuration.

Next, the feeding network is fabricated. The power divider is fabricated with equal magnitudes for the 12 output ports. To reduce the antenna fabrication cost, the phase shifters are fabricated with 2-bit phase quantization. Meanwhile, static phase shifters are used to ensure that the observed radiation performance is solely due to the pattern-reconfigurable antenna element, without interference from auxiliary tuning circuits. This simplification does not compromise the validity of the proposed ER-FAS proof-of-concept. The power divider is connected to the phase shifters through adapters, and the overall feeding network transmits the RF signal using 12 RF cables with equal length (40 cm).

Furthermore, the parasitic reflectors and directors are fabricated. Due to limited pump control (two ports per machine), the array uses silver electronic paste DM-SIP-3072S from Dycotec Materials. For proper metallization of the silver paste, a mask is prepared using Kapton film, and the metallic pattern is cut using a laser cutting machine. The silver paste is then screen-printed on a 125-$\mu$m-thick polyethylene terephthalate (PET) substrate. The overall fluid-metalized sample is cured in an oven at 70$^\circ$C for 2 hours. Finally, the directors, planar monopole antennas, and reflectors are assembled using 3D-printed supports, similar to those used for the array elements. The final fabricated ER-FAS prototype is shown in Fig.~\ref{fig_fabricated_FAS_Array}. It should be noted that the output port from the feeding network, without a phase shifter, acts as the $0^\circ$ phase reference.

\section{Experimental Results}\label{sec:Experiments}

\subsection{ER-FAS Element Measurement}

\begin{figure}[t]
    \centering
    % Left figure (a)
    \begin{minipage}[t]{0.4\linewidth}
        \centering
        \includegraphics[width=\linewidth]{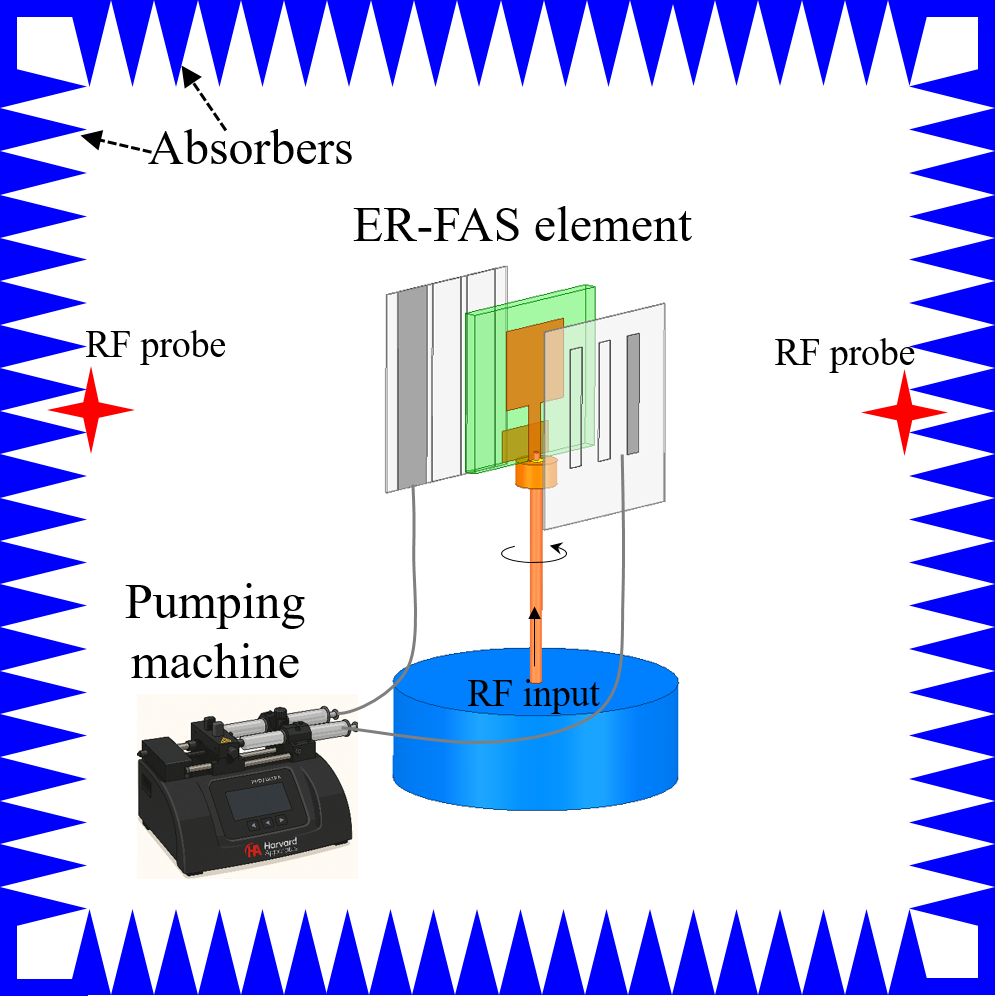}
        \footnotesize{{(a)}}
    \end{minipage}
    %\hfill
    \hspace{2mm}
    % Right figure (b)
    \begin{minipage}[t]{0.4\linewidth}
        \centering
        \includegraphics[width=\linewidth]{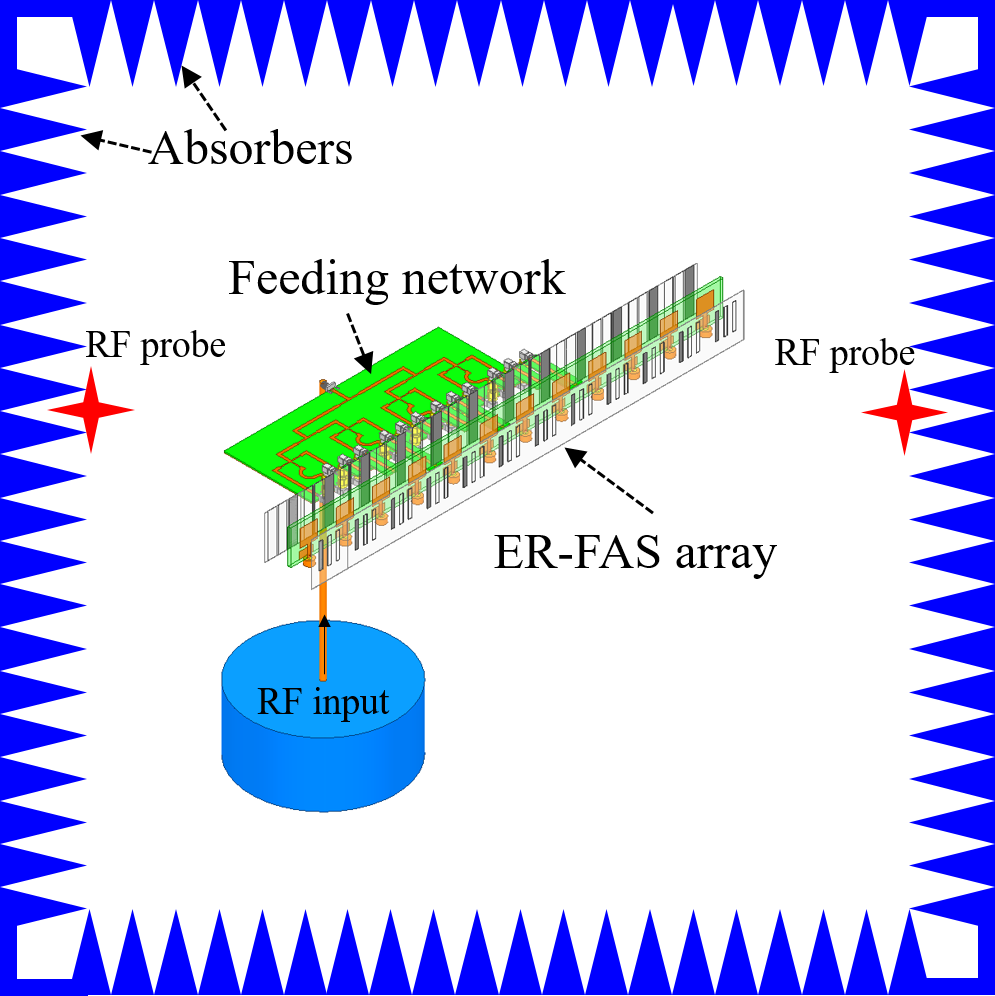}
        \footnotesize{{(b)}}
    \end{minipage}
    \caption{Schematic drawings of the test setups for the ER-FAS in anechoic chamber. (a) Element. (b) Array.}
    \label{fig_Revision_schematic_drawing}
\end{figure}

\begin{figure}[t]
  \centering
  \includegraphics[width=0.75\columnwidth]{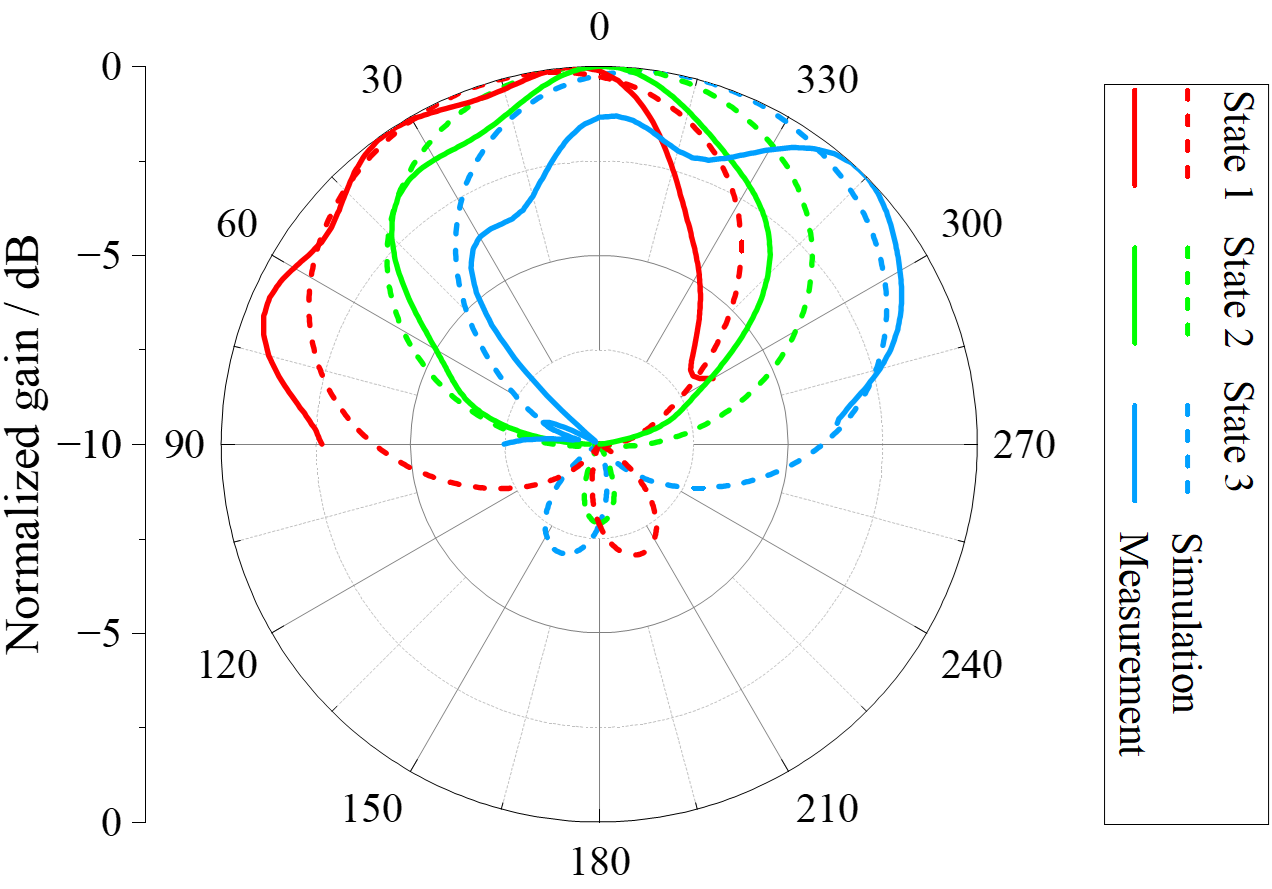}
  \caption{
    Comparison between simulation and measurement of the designed ER-FAS array element.
  }
  \label{fig_Sim_vs_Mea_Array_Element}
  \vspace{-1em}
\end{figure}

The schematic drawings of the full test setups for both the fabricated ER-FAS element and array in the anechoic chamber are shown in Fig.~\ref{fig_Revision_schematic_drawing}. The measurements were conducted in a Satimo chamber, which captures the radiated wave using near-field probes and transforms it into the far-field pattern through spherical wave expansion. The measured results, compared with the simulated results, are shown in Fig.~\ref{fig_Sim_vs_Mea_Array_Element}. From the normalized radiation patterns, it can be observed that the measured results generally follow the simulation trends, and three distinct patterns at different states can be obtained. The slight discrepancies are mainly attributed to fabrication errors of the channel structures, which can be further improved by using smoother and more precise channels. The detailed measurements of impedance matching performance, peak gain, and 3 dB beamwidth for more reconfigurable states have been summarized in Table~\ref{Element_Measurement_Table}.

\begin{table}[t]
  \renewcommand{\arraystretch}{1.3}
  \begin{center}
  \caption{Measurements of ER-FAS Element in Different States}\vspace{-0.5em}
  \label{Element_Measurement_Table}
  {
  \begin{tabular}{  c !{\color{black}\vrule width 1pt} c !{\color{black}\vrule} c !{\color{black}\vrule} c }
    \Xhline{1pt}
    \textbf{State} & $\bm{S_{11}}$ \textbf{(dB)} & \textbf{Gain (dBi)} & \textbf{3 dB Beamwidth} \\
    \Xhline{1pt}
    State 0   & $-27.1$ & $2.41$ & $174.0^\circ$ \\
    \hline
    State 1   & $-18.4$ & $4.04$ & $108.0^\circ$ \\
    \hline
    State 2   & $-13.3$ & $4.56$ & $89.5^\circ$  \\
    \hline
    State 3   & $-17.2$ & $4.32$ & $93.0^\circ$  \\
    \hline
    State 4   & $-18.2$ & $3.26$ & $80.0^\circ$  \\
    \hline
    State 63  & $-14.8$ & $5.19$ & $73.5^\circ$  \\
    \Xhline{1pt}
  \end{tabular}
  }
  \end{center}
\end{table}

\subsection{ER-FAS Array Measurement}

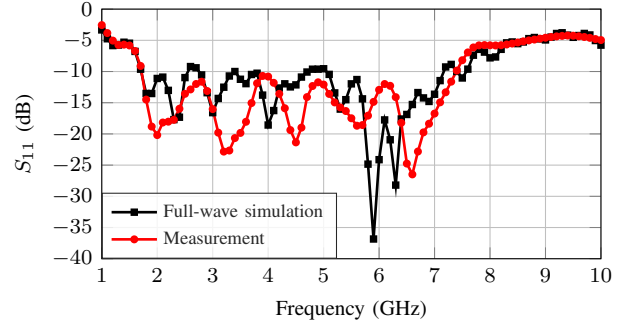
\begin{figure}[t]
    \centering
    % This file was created by matlab2tikz.
%
%The latest updates can be retrieved from
%  http://www.mathworks.com/matlabcentral/fileexchange/22022-matlab2tikz-matlab2tikz
%where you can also make suggestions and rate matlab2tikz.
%
\begin{tikzpicture}

\begin{axis}[%
width=2.6in,
height=1.3in,
at={(0in,0in)},
scale only axis,
xmin=1,
xmax=10,
xminorticks=true,
xtick={1,2,3,4,5,6,7,8,9,10},
xlabel style={font=\color{white!15!black},font=\footnotesize},
xticklabel style = {font=\color{white!15!black},font=\footnotesize},
xlabel={Frequency (GHz)},
ymin=-40,
ymax=0,
ytick={-40,-35,-30,-25,-20,-15,-10,-5,0},
ylabel style={font=\color{white!15!black},font=\footnotesize},
yticklabel style = {font=\color{white!15!black},font=\footnotesize},
ylabel={$S_{11}$ (dB)},
axis background/.style={fill=white},
xmajorgrids,
xminorgrids,
ymajorgrids,
legend style={at={(0,0)}, anchor=south west, font=\scriptsize, legend cell align=left, align=left, draw=white!15!black, fill opacity=0.85}
]
\addplot [color=black, line width=1pt, mark=square*, mark options={solid, black},mark size=0.9pt]
  table[row sep=crcr]{%
  1	-3.33111	\\
1.1	-4.74998	\\
1.2	-5.84263	\\
1.3	-5.75769	\\
1.4	-5.30658	\\
1.5	-5.46642	\\
1.6	-6.77769	\\
1.7	-9.62786	\\
1.8	-13.51745	\\
1.9	-13.52507	\\
2	-11.11253	\\
2.1	-10.898	\\
2.2	-13.01995	\\
2.3	-17.70781	\\
2.4	-17.3236	\\
2.5	-10.97383	\\
2.6	-9.20712	\\
2.7	-9.40498	\\
2.8	-10.54449	\\
2.9	-13.39178	\\
3	-16.59073	\\
3.1	-14.31489	\\
3.2	-12.51261	\\
3.3	-10.71688	\\
3.4	-10.00093	\\
3.5	-11.24111	\\
3.6	-11.94404	\\
3.7	-10.4779	\\
3.8	-10.27341	\\
3.9	-13.78027	\\
4	-18.57325	\\
4.1	-16.25699	\\
4.2	-12.6412	\\
4.3	-11.97034	\\
4.4	-12.46172	\\
4.5	-12.13871	\\
4.6	-10.89088	\\
4.7	-10.08059	\\
4.8	-9.62857	\\
4.9	-9.62319	\\
5	-9.55713	\\
5.1	-10.45938	\\
5.2	-13.42594	\\
5.3	-16.02091	\\
5.4	-14.51231	\\
5.5	-11.97698	\\
5.6	-11.28869	\\
5.7	-14.38155	\\
5.8	-24.8441	\\
5.9	-36.84053	\\
6	-24.12707	\\
6.1	-17.75386	\\
6.2	-20.93869	\\
6.3	-28.20158	\\
6.4	-17.61089	\\
6.5	-16.88302	\\
6.6	-15.2859	\\
6.7	-13.37426	\\
6.8	-14.22836	\\
6.9	-14.79604	\\
7	-13.64157	\\
7.1	-11.45797	\\
7.2	-9.27046	\\
7.3	-8.86187	\\
7.4	-9.9823	\\
7.5	-11.03455	\\
7.6	-9.62183	\\
7.7	-7.38549	\\
7.8	-6.44632	\\
7.9	-6.88857	\\
8	-7.8091	\\
8.1	-7.66494	\\
8.2	-6.44477	\\
8.3	-5.40863	\\
8.4	-5.26204	\\
8.5	-5.53054	\\
8.6	-5.26936	\\
8.7	-4.72927	\\
8.8	-4.54976	\\
8.9	-4.77015	\\
9	-4.93549	\\
9.1	-4.48184	\\
9.2	-3.90615	\\
9.3	-3.79171	\\
9.4	-4.16971	\\
9.5	-4.51762	\\
9.6	-4.23164	\\
9.7	-3.89858	\\
9.8	-4.13978	\\
9.9	-5.03423	\\
10	-5.77259	\\
};
\addlegendentry{Full-wave simulation}

\addplot [color=red, line width=1pt, mark=*, mark options={solid, red},mark size=1pt]
  table[row sep=crcr]{%
  1	-2.55893	\\
  1.1	-3.83739	\\
  1.2	-5.04675	\\
  1.3	-5.73883	\\
  1.4	-5.69275	\\
  1.5	-5.82488	\\
  1.6	-6.71431	\\
  1.7	-9.10732	\\
  1.8	-14.49791	\\
  1.9	-18.8227	\\
  2	-20.18763	\\
  2.1	-18.17184	\\
  2.2	-18.03727	\\
  2.3	-17.74052	\\
  2.4	-15.96435	\\
  2.5	-13.57457	\\
  2.6	-12.87207	\\
  2.7	-11.99989	\\
  2.8	-11.63441	\\
  2.9	-13.11917	\\
  3	-16.02329	\\
  3.1	-19.79482	\\
  3.2	-22.83237	\\
  3.3	-22.66181	\\
  3.4	-20.66372	\\
  3.5	-19.83988	\\
  3.6	-18.05593	\\
  3.7	-15.05104	\\
  3.8	-11.88714	\\
  3.9	-10.7098	\\
  4	-10.81057	\\
  4.1	-11.80966	\\
  4.2	-13.57683	\\
  4.3	-15.88425	\\
  4.4	-19.3831	\\
  4.5	-21.35619	\\
  4.6	-18.9916	\\
  4.7	-14.09848	\\
  4.8	-12.32319	\\
  4.9	-11.72893	\\
  5	-12.09033	\\
  5.1	-13.59224	\\
  5.2	-14.95205	\\
  5.3	-15.74546	\\
  5.4	-16.33055	\\
  5.5	-17.49137	\\
  5.6	-18.68648	\\
  5.7	-18.57803	\\
  5.8	-17.06729	\\
  5.9	-14.86241	\\
  6	-12.93723	\\
  6.1	-12.00305	\\
  6.2	-12.29169	\\
  6.3	-14.11232	\\
  6.4	-18.21852	\\
  6.5	-24.7522	\\
  6.6	-26.48552	\\
  6.7	-22.82014	\\
  6.8	-19.75012	\\
  6.9	-18.36522	\\
  7	-16.73814	\\
  7.1	-14.93871	\\
  7.2	-13.33153	\\
  7.3	-11.62678	\\
  7.4	-9.84971	\\
  7.5	-8.18914	\\
  7.6	-6.9396	\\
  7.7	-6.14099	\\
  7.8	-5.81026	\\
  7.9	-5.76534	\\
  8	-5.8015	\\
  8.1	-5.82558	\\
  8.2	-5.78329	\\
  8.3	-5.66781	\\
  8.4	-5.52049	\\
  8.5	-5.34614	\\
  8.6	-5.16545	\\
  8.7	-5.01348	\\
  8.8	-4.87897	\\
  8.9	-4.732	\\
  9	-4.58628	\\
  9.1	-4.46445	\\
  9.2	-4.37088	\\
  9.3	-4.28759	\\
  9.4	-4.23362	\\
  9.5	-4.25958	\\
  9.6	-4.35417	\\
  9.7	-4.46681	\\
  9.8	-4.60208	\\
  9.9	-4.79543	\\
  10	-4.95766	\\
};
\addlegendentry{Measurement}
\end{axis}

\end{tikzpicture}%
    \vspace{-3em}
    \caption{Simulated and measured S parameter of the designed ER-FAS.}
    \label{fig_S_parameter_Simulation_and_Measurement}
\end{figure}

The return loss performance is a crucial factor for FAS, which evaluates impedance matching. Therefore, the designed ER-FAS, including the planar monopole array, parasitic reflector, and director arrays, was measured using a PNA Network Analyzer (E8363C). The detailed measured S-parameter performance of the designed ER-FAS prototype, compared with the simulated results, is shown in Fig.~\ref{fig_S_parameter_Simulation_and_Measurement}. The measured results align well with the simulations. Specifically, the measured $S_{11}$ parameter is less than -10 dB from 1.8 GHz to 7.3 GHz, corresponding to a fractional bandwidth of 120.9\%. Therefore, the designed FAS exhibits wideband impedance matching performance, making it suitable for 5G sub-6 GHz applications.

\begin{figure}[t]
  \centering
  \begin{minipage}{\linewidth}
    \centering
    \resizebox{\linewidth}{!}{% This file was created by matlab2tikz.
%
%The latest updates can be retrieved from
%  http://www.mathworks.com/matlabcentral/fileexchange/22022-matlab2tikz-matlab2tikz
%where you can also make suggestions and rate matlab2tikz.
%
\begin{tikzpicture}[scale=0.8]

\begin{axis}[%
width= 0.8\linewidth,
height=0.3\linewidth,
scale only axis,
xmin=-90,
xmax=90,
xminorticks=true,
xtick={-90,-60,-30,0,30,60,90},
xlabel style={font=\color{white!15!black},font=\footnotesize,align=center},
xticklabel style = {font=\color{white!15!black},font=\footnotesize},
xlabel={ Azimuthal angle ($^\circ$) \\ (a) Beamforming to $\theta_\mathrm{az}=-90^\circ$},
ymin=-30,
ymax=1,
ytick={-30,-20,-10,0,10},
ylabel style={font=\color{white!15!black},font=\footnotesize},
yticklabel style = {font=\color{white!15!black},font=\footnotesize},
ylabel={Normalized gain (dB)},
axis background/.style={fill=white},
xmajorgrids,
xminorgrids,
ymajorgrids,
legend style={at={(1,1.1)}, anchor=south east, font=\scriptsize, legend cell align=left, align=left, draw=white!15!black, fill opacity=0.85}, legend columns=2
]
\addplot [color=blue, dashed, line width=1.2pt]
  table[row sep=crcr]{%
  -90	-5.61318	\\
-88	-5.2439	\\
-86	-4.85075	\\
-84	-4.44692	\\
-82	-4.04968	\\
-80	-3.68001	\\
-78	-3.36247	\\
-76	-3.12527	\\
-74	-3.00086	\\
-72	-3.02693	\\
-70	-3.24851	\\
-68	-3.72155	\\
-66	-4.51969	\\
-64	-5.74742	\\
-62	-7.56921	\\
-60	-10.28522	\\
-58	-14.58881	\\
-56	-23.05461	\\
-54	-27.92927	\\
-52	-18.64563	\\
-50	-15.7568	\\
-48	-15.27309	\\
-46	-16.65631	\\
-44	-20.28838	\\
-42	-26.886	\\
-40	-25.08804	\\
-38	-20.929	\\
-36	-20.03944	\\
-34	-21.88139	\\
-32	-25.89507	\\
-30	-24.10136	\\
-28	-20.00088	\\
-26	-18.30601	\\
-24	-18.50245	\\
-22	-19.53117	\\
-20	-19.37132	\\
-18	-18.1456	\\
-16	-17.84402	\\
-14	-19.25757	\\
-12	-21.57721	\\
-10	-20.88951	\\
-8	-18.86188	\\
-6	-18.90942	\\
-4	-22.32124	\\
-2	-30.11315	\\
0	-22.6948	\\
2	-18.89378	\\
4	-19.19959	\\
6	-24.49084	\\
8	-27.85979	\\
10	-18.53773	\\
12	-15.43121	\\
14	-15.57717	\\
16	-19.17722	\\
18	-24.79969	\\
20	-17.52947	\\
22	-13.50416	\\
24	-12.44316	\\
26	-13.83824	\\
28	-18.67886	\\
30	-26.6409	\\
32	-17.24194	\\
34	-13.2049	\\
36	-12.05169	\\
38	-12.9837	\\
40	-16.32742	\\
42	-23.65107	\\
44	-21.15109	\\
46	-15.39153	\\
48	-12.9536	\\
50	-12.42549	\\
52	-13.41662	\\
54	-15.75742	\\
56	-17.48149	\\
58	-14.68849	\\
60	-10.9661	\\
62	-8.21194	\\
64	-6.29731	\\
66	-4.98957	\\
68	-4.12754	\\
70	-3.60187	\\
72	-3.33443	\\
74	-3.26625	\\
76	-3.3507	\\
78	-3.54958	\\
80	-3.83087	\\
82	-4.16747	\\
84	-4.53652	\\
86	-4.91919	\\
88	-5.3006	\\
90	-5.66995	\\
};
\addlegendentry{Conventional antenna, simulation}

\addplot [color=red, dashed, line width=1.2pt]
  table[row sep=crcr]{%
  -90	-1.11386	\\
  -88	-0.83527	\\
  -86	-0.58154	\\
  -84	-0.35859	\\
  -82	-0.17669	\\
  -80	-0.05063	\\
  -78	-2.57E-06	\\
  -76	-0.0497	\\
  -74	-0.2309	\\
  -72	-0.5829	\\
  -70	-1.15635	\\
  -68	-2.01945	\\
  -66	-3.27016	\\
  -64	-5.06317	\\
  -62	-7.67944	\\
  -60	-11.75918	\\
  -58	-19.62042	\\
  -56	-27.98482	\\
  -54	-16.29905	\\
  -52	-12.76648	\\
  -50	-11.66132	\\
  -48	-12.21637	\\
  -46	-14.51907	\\
  -44	-19.42067	\\
  -42	-25.547	\\
  -40	-19.96604	\\
  -38	-16.818	\\
  -36	-16.62836	\\
  -34	-19.23068	\\
  -32	-25.05116	\\
  -30	-22.27459	\\
  -28	-17.75878	\\
  -26	-16.69887	\\
  -24	-18.95752	\\
  -22	-28.63013	\\
  -20	-24.37422	\\
  -18	-17.06474	\\
  -16	-15.14657	\\
  -14	-16.5587	\\
  -12	-23.13089	\\
  -10	-25.87283	\\
  -8	-16.92338	\\
  -6	-14.45824	\\
  -4	-15.43014	\\
  -2	-20.6587	\\
  0	-24.43235	\\
  2	-16.64525	\\
  4	-14.06242	\\
  6	-14.8129	\\
  8	-19.52277	\\
  10	-27.202	\\
  12	-18.01607	\\
  14	-14.80542	\\
  16	-15.08498	\\
  18	-18.99065	\\
  20	-35.04671	\\
  22	-21.81758	\\
  24	-16.86096	\\
  26	-16.10877	\\
  28	-18.22787	\\
  30	-23.30928	\\
  32	-23.32981	\\
  34	-18.90999	\\
  36	-17.26372	\\
  38	-17.77267	\\
  40	-19.47274	\\
  42	-19.50922	\\
  44	-17.12606	\\
  46	-15.19294	\\
  48	-14.54594	\\
  50	-15.24913	\\
  52	-17.37578	\\
  54	-20.26055	\\
  56	-19.28749	\\
  58	-15.3562	\\
  60	-12.26569	\\
  62	-10.1607	\\
  64	-8.75647	\\
  66	-7.84734	\\
  68	-7.29521	\\
  70	-7.00226	\\
  72	-6.89538	\\
  74	-6.91808	\\
  76	-7.02617	\\
  78	-7.18541	\\
  80	-7.36982	\\
  82	-7.56035	\\
  84	-7.74358	\\
  86	-7.91059	\\
  88	-8.05606	\\
  90	-8.1777	\\  
};
\addlegendentry{ER-FAS, simulation}

\addplot [color=blue, line width=1.2pt]
  table[row sep=crcr]{%
  -90	-4.4099	\\
-88	-3.83308	\\
-86	-3.25626	\\
-84	-2.67944	\\
-82	-2.30453	\\
-80	-2.05671	\\
-78	-1.93597	\\
-76	-2.12002	\\
-74	-2.49323	\\
-72	-3.05559	\\
-70	-4.13782	\\
-68	-5.64586	\\
-66	-7.57972	\\
-64	-11.36576	\\
-62	-16.65423	\\
-60	-23.44513	\\
-58	-16.46276	\\
-56	-11.70182	\\
-54	-9.16232	\\
-52	-8.94062	\\
-50	-9.80615	\\
-48	-11.7589	\\
-46	-17.49971	\\
-44	-19.38677	\\
-42	-17.42006	\\
-40	-14.75029	\\
-38	-13.89299	\\
-36	-14.84814	\\
-34	-15.61079	\\
-32	-14.33764	\\
-30	-11.02868	\\
-28	-9.52636	\\
-26	-9.096	\\
-24	-9.73762	\\
-22	-12.68062	\\
-20	-17.46048	\\
-18	-24.07718	\\
-16	-29.85473	\\
-14	-30.58478	\\
-12	-26.26733	\\
-10	-21.97854	\\
-8	-19.16128	\\
-6	-17.81554	\\
-4	-18.53937	\\
-2	-18.84916	\\
0	-18.74491	\\
2	-18.13407	\\
4	-17.61077	\\
6	-17.175	\\
8	-14.20168	\\
10	-11.64085	\\
12	-9.49251	\\
14	-9.33613	\\
16	-10.26062	\\
18	-12.26598	\\
20	-15.74745	\\
22	-17.32692	\\
24	-17.00438	\\
26	-18.02316	\\
28	-20.49384	\\
30	-24.41642	\\
32	-18.55065	\\
34	-14.71325	\\
36	-12.90425	\\
38	-14.29409	\\
40	-15.96157	\\
42	-17.9067	\\
44	-14.06508	\\
46	-11.08111	\\
48	-8.95483	\\
50	-8.68417	\\
52	-9.19161	\\
54	-10.47717	\\
56	-13.61227	\\
58	-16.03724	\\
60	-17.75209	\\
62	-13.32805	\\
64	-9.64647	\\
66	-6.70734	\\
68	-4.9371	\\
70	-3.55442	\\
72	-2.5593	\\
74	-2.14553	\\
76	-1.94356	\\
78	-1.95338	\\
80	-2.21667	\\
82	-2.52518	\\
84	-2.87892	\\
86	-3.23563	\\
88	-3.59234	\\
90	-3.94905	\\
};
\addlegendentry{Conventional antenna, measurement }

\addplot [color=red, line width=1.2pt]
  table[row sep=crcr]{%
  -90	-0.22568	\\
-88	-0.15046	\\
-86	-0.07523	\\
-84	0	\\
-82	-0.08694	\\
-80	-0.27393	\\
-78	-0.56096	\\
-76	-1.10635	\\
-74	-1.82929	\\
-72	-2.7298	\\
-70	-4.16899	\\
-68	-6.07643	\\
-66	-8.4521	\\
-64	-12.86944	\\
-62	-17.16614	\\
-60	-21.3422	\\
-58	-16.1464	\\
-56	-12.69519	\\
-54	-10.98857	\\
-52	-11.45876	\\
-50	-13.10906	\\
-48	-15.93947	\\
-46	-18.84217	\\
-44	-18.53865	\\
-42	-15.0289	\\
-40	-14.24651	\\
-38	-15.14989	\\
-36	-17.73903	\\
-34	-19.27063	\\
-32	-17.67478	\\
-30	-12.95146	\\
-28	-11.44766	\\
-26	-11.07956	\\
-24	-11.84718	\\
-22	-14.8036	\\
-20	-18.66325	\\
-18	-23.42613	\\
-16	-25.86403	\\
-14	-29.18241	\\
-12	-33.38128	\\
-10	-25.45006	\\
-8	-19.88113	\\
-6	-16.67449	\\
-4	-15.43071	\\
-2	-14.48197	\\
0	-13.82827	\\
2	-13.63023	\\
4	-13.7486	\\
6	-14.18339	\\
8	-13.85954	\\
10	-12.93041	\\
12	-11.39601	\\
14	-10.63955	\\
16	-10.62295	\\
18	-11.34619	\\
20	-13.55916	\\
22	-15.95697	\\
24	-18.53963	\\
26	-19.67349	\\
28	-19.92595	\\
30	-19.29701	\\
32	-16.49703	\\
34	-14.30484	\\
36	-12.72045	\\
38	-13.02902	\\
40	-13.78572	\\
42	-14.99056	\\
44	-13.26447	\\
46	-11.25735	\\
48	-8.9692	\\
50	-8.01091	\\
52	-7.66021	\\
54	-7.91709	\\
56	-9.28088	\\
58	-11.24723	\\
60	-13.81613	\\
62	-17.18445	\\
64	-18.16187	\\
66	-16.74836	\\
68	-13.68448	\\
70	-11.17956	\\
72	-9.23362	\\
74	-8.15061	\\
76	-7.40589	\\
78	-6.99947	\\
80	-7.12952	\\
82	-7.42943	\\
84	-7.89917	\\
86	-8.41677	\\
88	-8.93437	\\
90	-9.45197	\\
};
\addlegendentry{ER-FAS, measurement}

\end{axis}

\end{tikzpicture}%
}
  \end{minipage}

  \vspace{0.8em}

  \begin{minipage}{\linewidth}
    \centering
    \resizebox{\linewidth}{!}{% This file was created by matlab2tikz.
%
%The latest updates can be retrieved from
%  http://www.mathworks.com/matlabcentral/fileexchange/22022-matlab2tikz-matlab2tikz
%where you can also make suggestions and rate matlab2tikz.
%

\begin{tikzpicture}[scale=0.8]

\begin{axis}[%
width=0.8\linewidth,
height=0.3\linewidth,
scale only axis,
xmin=-90,
xmax=90,
xminorticks=true,
xtick={-90,-60,-30,0,30,60,90},
xlabel style={font=\color{white!15!black},font=\footnotesize,align=center},
xticklabel style = {font=\color{white!15!black},font=\footnotesize},
xlabel={Azimuthal angle ($^\circ$)  \\ (b) Beamforming to $\theta_\mathrm{az}=60^\circ$},
ymin=-30,
ymax=1,
ytick={-30,-20,-10,0,10},
ylabel style={font=\color{white!15!black},font=\footnotesize},
yticklabel style = {font=\color{white!15!black},font=\footnotesize},
ylabel={Normalized gain (dB)},
axis background/.style={fill=white},
xmajorgrids,
xminorgrids,
ymajorgrids,
legend style={at={(1,1.1)}, anchor=south east, font=\scriptsize, legend cell align=left, align=left, draw=white!15!black, fill opacity=0.85,legend columns=2}
]
\addplot [color=blue, dashed, line width=1.2pt]
  table[row sep=crcr]{%
  -90	-11.97819	\\
-88	-11.82094	\\
-86	-11.80196	\\
-84	-11.95536	\\
-82	-12.33117	\\
-80	-13.00379	\\
-78	-14.09074	\\
-76	-15.79864	\\
-74	-18.56106	\\
-72	-23.63936	\\
-70	-40.44486	\\
-68	-25.12278	\\
-66	-18.72303	\\
-64	-15.39968	\\
-62	-13.49175	\\
-60	-12.54995	\\
-58	-12.41787	\\
-56	-13.00712	\\
-54	-14.05407	\\
-52	-14.63843	\\
-50	-13.63574	\\
-48	-11.71056	\\
-46	-10.05568	\\
-44	-9.10692	\\
-42	-8.94529	\\
-40	-9.57022	\\
-38	-10.93902	\\
-36	-12.90121	\\
-34	-15.13292	\\
-32	-17.44755	\\
-30	-20.60361	\\
-28	-27.31866	\\
-26	-29.68842	\\
-24	-20.92154	\\
-22	-17.51862	\\
-20	-16.64485	\\
-18	-17.68857	\\
-16	-20.64667	\\
-14	-25.98105	\\
-12	-31.81588	\\
-10	-24.60526	\\
-8	-18.47455	\\
-6	-14.621	\\
-4	-12.55302	\\
-2	-12.14664	\\
0	-13.57716	\\
2	-17.60227	\\
4	-26.07954	\\
6	-22.67346	\\
8	-18.75027	\\
10	-18.59723	\\
12	-19.38951	\\
14	-17.07678	\\
16	-14.15538	\\
18	-12.87722	\\
20	-13.37527	\\
22	-15.28649	\\
24	-16.3352	\\
26	-14.42904	\\
28	-12.63771	\\
30	-12.40902	\\
32	-14.01236	\\
34	-17.60125	\\
36	-19.77222	\\
38	-16.43744	\\
40	-14.21799	\\
42	-14.32606	\\
44	-17.51937	\\
46	-31.01958	\\
48	-17.47326	\\
50	-10.3505	\\
52	-6.48121	\\
54	-4.04901	\\
56	-2.48858	\\
58	-1.53831	\\
60	-1.04911	\\
62	-0.92377	\\
64	-1.0922	\\
66	-1.49974	\\
68	-2.10067	\\
70	-2.85444	\\
72	-3.72314	\\
74	-4.66983	\\
76	-5.65744	\\
78	-6.64835	\\
80	-7.60473	\\
82	-8.48985	\\
84	-9.2704	\\
86	-9.91937	\\
88	-10.41897	\\
90	-10.7624	\\
};
\addlegendentry{Conventional antenna, simulation}

\addplot [color=red, dashed, line width=1.2pt]
  table[row sep=crcr]{%
  -90	-17.07287	\\
-88	-17.21851	\\
-86	-17.51702	\\
-84	-17.99517	\\
-82	-18.7044	\\
-80	-19.73495	\\
-78	-21.24862	\\
-76	-23.56394	\\
-74	-27.41739	\\
-72	-34.12643	\\
-70	-30.66961	\\
-68	-24.18778	\\
-66	-20.2995	\\
-64	-17.80453	\\
-62	-16.2788	\\
-60	-15.61653	\\
-58	-15.90546	\\
-56	-17.49234	\\
-54	-21.40717	\\
-52	-28.93202	\\
-50	-20.60925	\\
-48	-14.62269	\\
-46	-11.24545	\\
-44	-9.29857	\\
-42	-8.36351	\\
-40	-8.22642	\\
-38	-8.68523	\\
-36	-9.44495	\\
-34	-10.19149	\\
-32	-10.95337	\\
-30	-12.29636	\\
-28	-15.23521	\\
-26	-22.28323	\\
-24	-25.62021	\\
-22	-16.79205	\\
-20	-13.79279	\\
-18	-13.38234	\\
-16	-14.79729	\\
-14	-17.29929	\\
-12	-19.31346	\\
-10	-21.17129	\\
-8	-24.55682	\\
-6	-19.26148	\\
-4	-13.44701	\\
-2	-10.49634	\\
0	-9.72194	\\
2	-11.11025	\\
4	-15.72466	\\
6	-28.9565	\\
8	-17.11267	\\
10	-12.6244	\\
12	-11.62411	\\
14	-12.77499	\\
16	-15.53418	\\
18	-18.00728	\\
20	-17.60226	\\
22	-16.56385	\\
24	-15.84297	\\
26	-14.68268	\\
28	-13.30365	\\
30	-12.59976	\\
32	-13.09551	\\
34	-15.06481	\\
36	-18.3739	\\
38	-19.92281	\\
40	-17.85217	\\
42	-16.455	\\
44	-16.3014	\\
46	-15.34373	\\
48	-11.97473	\\
50	-8.27436	\\
52	-5.34745	\\
54	-3.21067	\\
56	-1.71897	\\
58	-0.74607	\\
60	-0.19699	\\
62	0	\\
64	-0.09903	\\
66	-0.44831	\\
68	-1.0087	\\
70	-1.7451	\\
72	-2.62441	\\
74	-3.61384	\\
76	-4.67949	\\
78	-5.78508	\\
80	-6.89113	\\
82	-7.9548	\\
84	-8.93098	\\
86	-9.77499	\\
88	-10.4471	\\
90	-10.91824	\\
};
\addlegendentry{ER-FAS, simulation}

\addplot [color=blue, line width=1.2pt]
  table[row sep=crcr]{%
  -90	-15.9927	\\
  -88	-16.16877	\\
  -86	-16.98439	\\
  -84	-17.92363	\\
  -82	-18.98647	\\
  -80	-17.9094	\\
  -78	-16.0871	\\
  -76	-13.51956	\\
  -74	-11.43914	\\
  -72	-9.67972	\\
  -70	-8.24131	\\
  -68	-7.35538	\\
  -66	-6.76758	\\
  -64	-6.47791	\\
  -62	-6.99038	\\
  -60	-8.07981	\\
  -58	-9.7462	\\
  -56	-12.26085	\\
  -54	-13.00744	\\
  -52	-11.98598	\\
  -50	-9.94189	\\
  -48	-8.7085	\\
  -46	-8.28582	\\
  -44	-9.18807	\\
  -42	-10.04099	\\
  -40	-10.84458	\\
  -38	-10.27745	\\
  -36	-9.8773	\\
  -34	-9.64413	\\
  -32	-10.79617	\\
  -30	-12.90143	\\
  -28	-15.95991	\\
  -26	-20.34051	\\
  -24	-20.39716	\\
  -22	-16.12985	\\
  -20	-13.45544	\\
  -18	-11.89318	\\
  -16	-11.4431	\\
  -14	-12.06499	\\
  -12	-11.4696	\\
  -10	-9.65691	\\
  -8	-7.19644	\\
  -6	-5.46827	\\
  -4	-4.47241	\\
  -2	-4.97999	\\
  0	-6.159	\\
  2	-8.00946	\\
  4	-10.56706	\\
  6	-13.10908	\\
  8	-15.63551	\\
  10	-20.14757	\\
  12	-20.85708	\\
  14	-17.76401	\\
  16	-14.48365	\\
  18	-12.64185	\\
  20	-12.23862	\\
  22	-13.21756	\\
  24	-13.91691	\\
  26	-14.33666	\\
  28	-15.19144	\\
  30	-18.16089	\\
  32	-23.24503	\\
  34	-19.42334	\\
  36	-15.58369	\\
  38	-11.72608	\\
  40	-11.63217	\\
  42	-13.955	\\
  44	-18.69459	\\
  46	-16.93963	\\
  48	-13.06852	\\
  50	-7.08124	\\
  52	-4.61064	\\
  54	-2.80956	\\
  56	-1.67798	\\
  58	-1.26519	\\
  60	-1.06627	\\
  62	-1.08119	\\
  64	-1.40432	\\
  66	-1.88114	\\
  68	-2.51166	\\
  70	-3.48988	\\
  72	-4.62461	\\
  74	-5.91584	\\
  76	-7.26037	\\
  78	-8.3792	\\
  80	-9.27233	\\
  82	-9.43229	\\
  84	-9.59225	\\
  86	-9.75221	\\
  88	-10.91217	\\
  90	-11.07213	\\
};
\addlegendentry{Conventional antenna, measurement}

\addplot [color=red, line width=1.2pt]
  table[row sep=crcr]{%
  -90	-21.96389	\\
  -88	-20.16667	\\
  -86	-18.71867	\\
  -84	-17.49307	\\
  -82	-16.48986	\\
  -80	-15.5834	\\
  -78	-14.66735	\\
  -76	-13.74171	\\
  -74	-12.86674	\\
  -72	-12.03943	\\
  -70	-11.25977	\\
  -68	-10.64571	\\
  -66	-10.23097	\\
  -64	-10.01553	\\
  -62	-10.66064	\\
  -60	-12.14926	\\
  -58	-14.48141	\\
  -56	-17.466	\\
  -54	-16.77086	\\
  -52	-12.39596	\\
  -50	-9.78547	\\
  -48	-8.09868	\\
  -46	-7.33558	\\
  -44	-7.62144	\\
  -42	-7.94566	\\
  -40	-8.30822	\\
  -38	-8.33002	\\
  -36	-8.62089	\\
  -34	-9.18083	\\
  -32	-11.1738	\\
  -30	-14.4585	\\
  -28	-19.03493	\\
  -26	-23.68781	\\
  -24	-22.93155	\\
  -22	-16.76614	\\
  -20	-14.06171	\\
  -18	-12.21001	\\
  -16	-11.21105	\\
  -14	-11.16136	\\
  -12	-10.76215	\\
  -10	-10.01344	\\
  -8	-8.56396	\\
  -6	-7.57351	\\
  -4	-7.0421	\\
  -2	-7.99153	\\
  0	-9.62748	\\
  2	-11.94997	\\
  4	-13.72663	\\
  6	-14.88592	\\
  8	-15.42784	\\
  10	-17.16113	\\
  12	-17.5472	\\
  14	-16.58604	\\
  16	-14.63161	\\
  18	-13.81336	\\
  20	-14.13129	\\
  22	-15.84767	\\
  24	-16.43661	\\
  26	-15.89811	\\
  28	-15.66769	\\
  30	-17.44102	\\
  32	-21.21813	\\
  34	-21.41399	\\
  36	-19.06345	\\
  38	-14.16652	\\
  40	-14.00173	\\
  42	-16.77944	\\
  44	-22.49966	\\
  46	-17.07134	\\
  48	-11.78341	\\
  50	-6.63588	\\
  52	-4.21902	\\
  54	-2.38891	\\
  56	-1.14555	\\
  58	-0.55733	\\
  60	-0.17548	\\
  62	0	\\
  64	-0.14803	\\
  66	-0.46501	\\
  68	-0.95094	\\
  70	-1.82789	\\
  72	-2.90565	\\
  74	-4.18421	\\
  76	-5.73883	\\
  78	-7.2643	\\
  80	-8.76064	\\
  82	-9.78231	\\
  84	-10.80398	\\
  86	-11.82565	\\
  88	-12.84732	\\
  90	-13.86899	\\
};
\addlegendentry{ER-FAS, measurement}

\end{axis}

\end{tikzpicture}%
}
  \end{minipage}
  \caption{Measured radiation pattern compared with full-wave simulation for array beamforming at
  $\theta_\mathrm{az}=-90^\circ$ and $\theta_\mathrm{az}=60^\circ$.}
  \label{fig_patterns_sim_mea}
\end{figure}
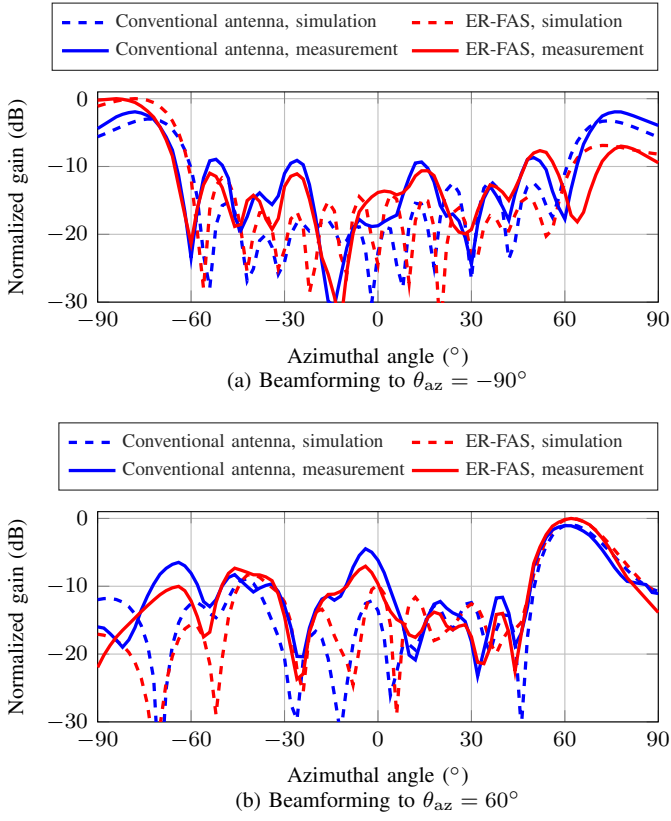

To experimentally verify the simulation results for the far-field gain enhancement of the designed ER-FAS, as shown in Fig.~\ref{fig_RPs_m90} and Fig.~\ref{fig_RPs_60}, the entire ER-FAS prototype was measured in the Satimo anechoic chamber, as demonstrated in Fig.~\ref{fig_fabricated_FAS_Array}. The deployed Satimo anechoic chamber measures the near field of the antenna and then converts it to the far-field radiation pattern using a post-processing algorithm. The measured far-field normalized radiation pattern, compared with the simulation, for the $-90^\circ$ beamforming is shown in Fig.~\ref{fig_patterns_sim_mea}-(a). The measurement results closely match the simulated radiation patterns for both the conventional and electromagnetically reconfigurable antenna arrays. At the large desired beamforming angle of $-90^\circ$ (end-fire radiation), the measured result shows a gain enhancement of 4.2 dB for the designed ER-FAS compared to the conventional antenna array, which is consistent with the simulated enhancement of 4.5 dB. For the beamforming towards $60^\circ$, the simulated and measured far-field radiation patterns are shown in Fig.~\ref{fig_patterns_sim_mea}-(b). The experimental results indicate a gain enhancement of approximately 1.1 dB at the desired array beam of $60^\circ$, which matches the simulated enhancement of 0.9 dB. It should be noted that the measurement was conducted with 2-bit quantization for the phase shifters. The proposed ER-FAS balances antenna performance and fluidic control complexity in both hardware and algorithm. The hardware design uses only parasitic strip-type directors and reflectors without complex geometry, leading to low control complexity. Meanwhile, the joint beamforming algorithm exhibits low computational burden. As a result, additional electromagnetic degrees of freedom are achieved with minimized overhead in both hardware and algorithm. 

\begin{figure}[t]
\centering
\includegraphics[width=0.8\linewidth]{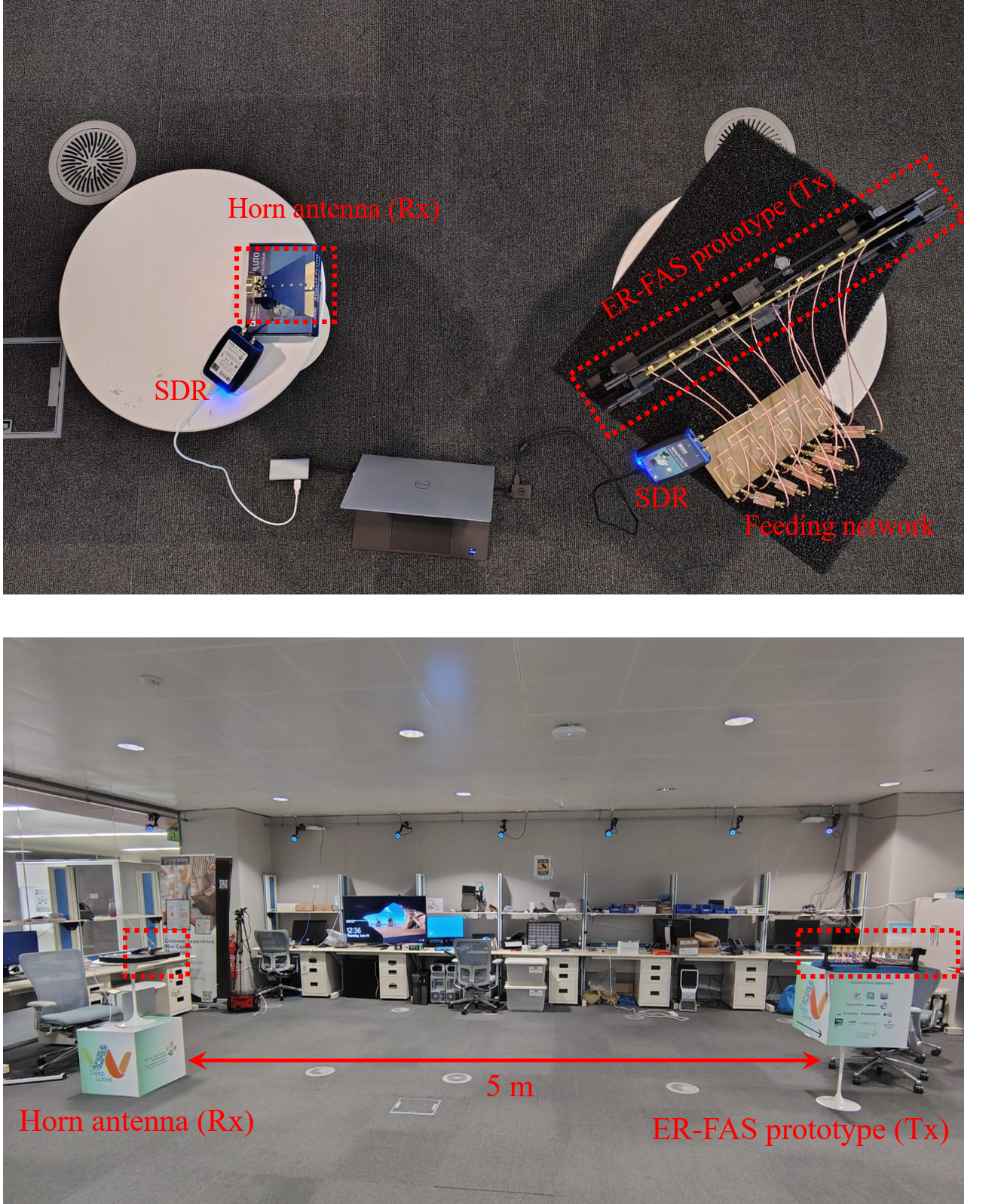}
\vspace{-0.5em}
\caption{Experimental setup at various distances for SDR-based validation.}
\label{fig:SDRExpeSetup}
\end{figure}

\begin{figure*}[t]
    \centering
    % (a)
    \begin{minipage}{0.44\textwidth}
        \centering
        \begin{tikzpicture}
            \begin{axis}[
                width=\linewidth,
                height=1.7in,
                ybar,
                ymin=-50, ymax=0,
                ytick={-50,-40,-30,-20,-10,0},
                symbolic x coords={0.3m, 0.6m, 0.9m, 1.2m},
                xtick=data,
                xticklabels={$d_\mathrm{TR}=0.3\,\mathrm{m}$, $d_\mathrm{TR}=0.6\,\mathrm{m}$, $d_\mathrm{TR}=0.9\,\mathrm{m}$, $d_\mathrm{TR}=1.2\,\mathrm{m}$},
                xlabel style={font=\footnotesize},
                xticklabel style = {font=\tiny},
                ylabel={Received power (dBm)},
                ylabel style={font=\footnotesize},
                yticklabel style = {font=\footnotesize},
                nodes near coords,
                every node near coord/.append style={font=\tiny},
                enlarge x limits=0.2,
                bar width=17pt,
                ymajorgrids=true,
                grid style=dashed,
                legend style={at={(0.97,0.93)}, anchor=north east, font=\footnotesize}
            ]
                \addplot [fill=blue!60] coordinates {(0.3m,-30.054) (0.6m,-35.372) (0.9m,-34.889) (1.2m,-42.474)};
                \addplot [fill=red!60] coordinates {(0.3m,-25.575) (0.6m,-29.141) (0.9m,-32.014) (1.2m,-35.092)};
                \legend{Conventional antenna, ER-FAS}
            \end{axis}
        \end{tikzpicture}
        \\ \footnotesize (a)
    \end{minipage}%
    \begin{minipage}{0.28\textwidth}
        \centering
        \begin{tikzpicture}
            \begin{axis}[
                width=\linewidth,
                height=1.6in,
                xmin=0.2000, xmax=0.5120,
                ymin=0, ymax=0.06,
                xticklabel style = {font=\footnotesize},
                xlabel={Transmit power (mW)},
                xlabel style={font=\footnotesize},
                yticklabel style = {font=\footnotesize},
                ylabel={Bit error rate},
                ylabel style={font=\footnotesize},
                xmajorgrids, xminorgrids, ymajorgrids,
                legend style={at={(1,1)}, anchor=north east, font=\scriptsize}
            ]
                \addplot [color=blue, line width=1.2pt, mark=diamond, mark size=2.5pt]
                table[row sep=crcr]{%
0.200000000000000	0.0575305486332192\\
0.220500000000000	0.0427881838706290\\
0.242000000000000	0.0283539656426352\\
0.264500000000000	0.0212745522393794\\
0.288000000000000	0.0120680033083711\\
0.312500000000000	0.00872517984253381\\
0.338000000000000	0.00465602497560607\\
0.364500000000000	0.00316607810540300\\
0.392000000000000	0.00278601100448377\\
0.420500000000000	0.00200293111871031\\
0.450000000000000	0.00161807086935723\\
0.480500000000000	0.00121133629691857\\
0.512000000000000	0.00106829080471840\\
};
                \addlegendentry{Conventional antenna}

                \addplot [color=red, line width=1.2pt, mark=o]
                table[row sep=crcr]{%
0.200000000000000	0.0296785900958885\\
0.220500000000000	0.0189892347746110\\
0.242000000000000	0.0124545586374842\\
0.264500000000000	0.00663393888480521\\
0.288000000000000	0.00463299380447338\\
0.312500000000000	0.00263230239738917\\
0.338000000000000	0.00185990002698616\\
0.364500000000000	0.00150536933570388\\
0.392000000000000	0.00106525721352982\\
0.420500000000000	0.00107360212574591\\
0.450000000000000	0.000787679856013568\\
0.480500000000000	0.000931233000958928\\
0.512000000000000	0.000471695266511447\\
};
                \addlegendentry{ER-FAS}
            \end{axis}
        \end{tikzpicture}
        \\ \footnotesize (b)
    \end{minipage}%
    \begin{minipage}{0.28\textwidth}
        \centering
        \begin{tikzpicture}
            \begin{axis}[
                width=\linewidth,
                height=1.6in,
                xmin=0.4500, xmax=2.00,
                ymin=0, ymax=0.24,
                xticklabel style = {font=\footnotesize},
                xlabel={Transmit power (mW)},
                xlabel style={font=\footnotesize},
                yticklabel style = {font=\footnotesize},
                ylabel={Bit error rate},
                ylabel style={font=\footnotesize},
                xmajorgrids, xminorgrids, ymajorgrids,
                legend style={at={(1,1)}, anchor=north east, font=\scriptsize}
            ]
                \addplot [color=blue, line width=1.2pt, mark=diamond, mark size=2.5pt]
                table[row sep=crcr]{%
0.450000000000000	NaN\\
0.612500000000000	0.121835556022262\\
0.800000000000000	0.0683345584374672\\
1.01250000000000	0.0166159123525500\\
1.25000000000000	0.00594350519794183\\
1.51250000000000	0.00227986045480533\\
1.80000000000000	0.000739729543677879\\
2.11250000000000	0.00116034862963352\\
2.45000000000000	0.000527669851937415\\
};
                \addlegendentry{Conventional antenna}

                \addplot [color=red, line width=1.2pt, mark=o]
                table[row sep=crcr]{%
0.450000000000000	0.0845998576545673\\
0.612500000000000	0.0368875354405124\\
0.800000000000000	0.0101984668696839\\
1.01250000000000	0.00461199201932164\\
1.25000000000000	0.000976583009555812\\
1.51250000000000	0.000873674262312297\\
1.80000000000000	0.00146592460359131\\
2.11250000000000	0.000805943505197942\\
2.45000000000000	0.000287024397073751\\
};
                \addlegendentry{ER-FAS}
            \end{axis}
        \end{tikzpicture}
        \\ \footnotesize (c)
    \end{minipage}

    \vspace{-0.3em}
    \caption{Illustration of the SDR communication experiment results at $\theta_\mathrm{az} = -90^\circ$. (a) Received power measurement under various Tx-Rx distances. (b) BER measurements at $d_\mathrm{TR}=\unit[1.2]{m}$. (c) BER measurements at $d_\mathrm{TR}=\unit[5]{m}$.}
    \label{fig:com90}
\end{figure*}
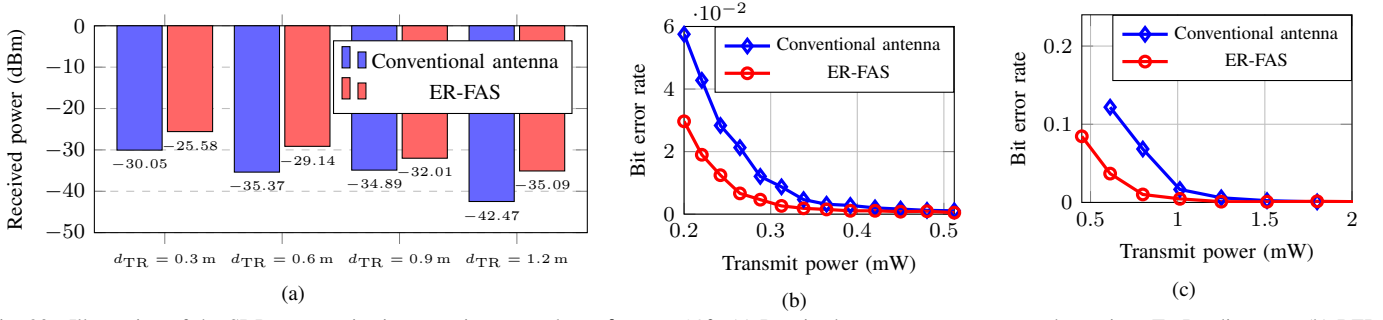

\begin{figure*}[t]
    \centering
    % (a)
    \begin{minipage}{0.44\textwidth}
        \centering
        \begin{tikzpicture}
            \begin{axis}[
                width=\linewidth,
                height=1.7in,
                ybar,
                ymin=-50, ymax=0,
                ytick={-50,-40,-30,-20,-10,0},
                symbolic x coords={0.3m, 0.6m, 0.9m, 1.2m},
                xtick=data,
                xticklabels={$d_\mathrm{TR}=0.3\,\mathrm{m}$, $d_\mathrm{TR}=0.6\,\mathrm{m}$, $d_\mathrm{TR}=0.9\,\mathrm{m}$, $d_\mathrm{TR}=1.2\,\mathrm{m}$},
                xlabel style={font=\footnotesize},
                xticklabel style = {font=\tiny},
                ylabel={Received power (dBm)},
                ylabel style={font=\footnotesize},
                yticklabel style = {font=\footnotesize},
                nodes near coords,
                every node near coord/.append style={font=\tiny},
                enlarge x limits=0.2,
                bar width=17pt,
                ymajorgrids=true,
                grid style=dashed,
                legend style={at={(0.97,0.93)}, anchor=north east, font=\footnotesize}
            ]
                % Blue bars (Conventional Antenna)
                \addplot [fill=blue!60] coordinates {(0.3m,-30.546) (0.6m,-33.680) (0.9m,-34.335) (1.2m,-39.754)};
                % Orange bars (Element-Reconfigurable Array)
                \addplot [fill=red!60] coordinates {(0.3m,-29.747) 
                (0.6m,-32.402) (0.9m,-33.898) (1.2m,-38.308)};

                \legend{Conventional antenna, ER-FAS}
            \end{axis}
        \end{tikzpicture}
        \\ \footnotesize (a)
    \end{minipage}%
    \begin{minipage}{0.28\textwidth}
        \centering
        \begin{tikzpicture}
            \begin{axis}[
                width=\linewidth,
                height=1.6in,
                xmin=0.2000, xmax=0.5120,
                ymin=0, ymax=0.06,
                xticklabel style = {font=\footnotesize},
                xlabel={Transmit power (mW)},
                xlabel style={font=\footnotesize},
                yticklabel style = {font=\footnotesize},
                ylabel={Bit error rate},
                ylabel style={font=\footnotesize},
                xmajorgrids, xminorgrids, ymajorgrids,
                legend style={at={(1,1)}, anchor=north east, font=\scriptsize}
            ]
                \addplot [color=blue, line width=1.2pt, mark=diamond, mark size=2.5pt]
                table[row sep=crcr]{%
0.200000000000000	0.0501776861559548\\
0.220500000000000	0.0501782234823293\\
0.242000000000000	0.0337140702414251\\
0.264500000000000	0.0228254813423847\\
0.288000000000000	0.0188601347570240\\
0.312500000000000	0.0118601600123352\\
0.338000000000000	0.0115617213826812\\
0.364500000000000	0.00658823469874457\\
0.392000000000000	0.00452069460386892\\
0.420500000000000	0.00137094986407178\\
0.450000000000000	0.00138234221818967\\
0.480500000000000	0.00107911419661682\\
0.512000000000000	0.00113829675522419\\
};
                \addlegendentry{Conventional antenna}

                \addplot [color=red, line width=1.2pt, mark=o]
                table[row sep=crcr]{%
0.200000000000000	0.0388462175911459\\
0.220500000000000	0.0258250179113326\\
0.242000000000000	0.0181582471547119\\
0.264500000000000	0.0105370246793558\\
0.288000000000000	0.00605726486751374\\
0.312500000000000	0.00304847088473108\\
0.338000000000000	0.00315232229522118\\
0.364500000000000	0.00151879968620862\\
0.392000000000000	0.00134704473136031\\
0.420500000000000	0.00104436149800006\\
0.450000000000000	0.000880165723177379\\
0.480500000000000	0.000730953435389834\\
0.512000000000000	0.000715240570729411\\
};
                \addlegendentry{ER-FAS}
            \end{axis}
        \end{tikzpicture}
        \\ \footnotesize (b)
    \end{minipage}%
    \begin{minipage}{0.28\textwidth}
        \centering
        \begin{tikzpicture}
            \begin{axis}[
                width=\linewidth,
                height=1.6in,
                xmin=0.4500, xmax=4.000,
                ymin=0, ymax=0.3,
                xticklabel style = {font=\footnotesize},
                xlabel={Transmit power (mW)},
                xlabel style={font=\footnotesize},
                yticklabel style = {font=\footnotesize},
                ylabel={Bit error rate},
                ylabel style={font=\footnotesize},
                xmajorgrids, xminorgrids, ymajorgrids,
                legend style={at={(1,1)}, anchor=north east, font=\scriptsize}
            ]
                \addplot [color=blue, line width=1.2pt, mark=diamond, mark size=2.5pt]
                table[row sep=crcr]{%
0.450000000000000	NaN\\
0.612500000000000	NaN\\
0.800000000000000	NaN\\
1.01250000000000	NaN\\
1.25000000000000	0.144046743973238\\
1.51250000000000	0.0950145180945032\\
1.80000000000000	0.0765477373269036\\
2.11250000000000	0.0458768245300851\\
2.45000000000000	0.0273878645538529\\
2.81250000000000	0.00691472108462553\\
3.20000000000000	0.00199366946190426\\
3.61250000000000	0.00207429259394025\\
4.05000000000000    0.00247534676811165\\
};
                \addlegendentry{Conventional antenna}

                \addplot [color=red, line width=1.2pt, mark=o]
                table[row sep=crcr]{%
0.450000000000000	NaN\\
0.612500000000000	0.153873079211733\\
0.800000000000000	0.0935332430085826\\
1.01250000000000	0.0322294500295683\\
1.25000000000000	0.0152472960201617\\
1.51250000000000	0.00606601561132696\\
1.80000000000000	0.00824002598360396\\
2.11250000000000	0.00337503155853354\\
2.45000000000000	0.00120617070632822\\
2.81250000000000	0.000589050069255887\\
3.20000000000000	0.000910891379255609\\
3.61250000000000	0.000336961975101217\\
4.05000000000000    0.000292158166777509\\
};
                \addlegendentry{ER-FAS}
            \end{axis}
        \end{tikzpicture}
        \\ \footnotesize (c)
    \end{minipage}

    \vspace{-0.3em}
    \caption{Illustration of the SDR communication experiment results at $\theta_\mathrm{az} = 60^\circ$. (a) Received power measurement under various Tx-Rx distances. (b) BER measurements at $d_\mathrm{TR}=\unit[1.2]{m}$. (c) BER measurements at $d_\mathrm{TR}=\unit[5]{m}$.}
    \label{fig:com60}
\end{figure*}
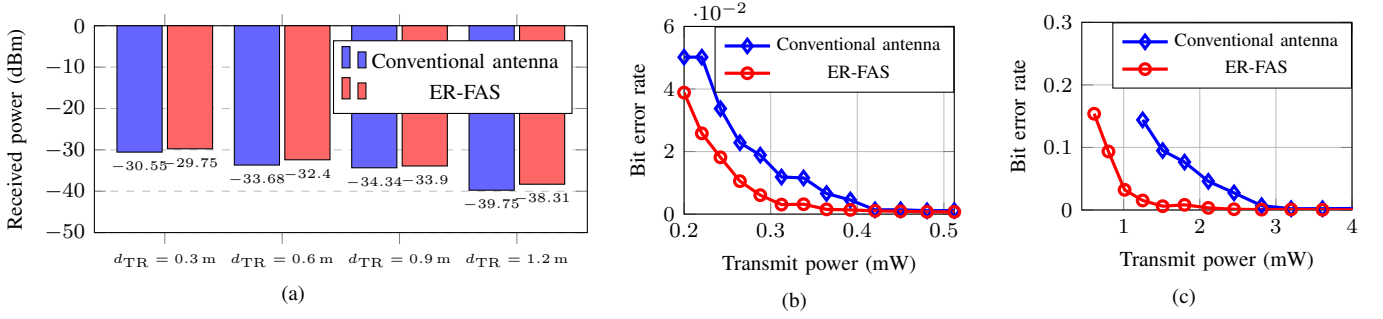

\subsection{SDR-based Communication Experiment}
Last but not least, a series of indoor experiments are conducted using the Analog Devices ADALM-Pluto SDR, a versatile SDR platform designed for prototyping various digital communication systems in real time. It operates over a very wide frequency range, i.e., 325 MHz to 3.8 GHz, and could support an instantaneous bandwidth of up to 20 MHz. 

In the communication trails, two ADALM-Pluto SDRs are utilized to investigate the performance of a point-to-point communication link whereby the transmit SDR sends data using the ER-FAS antenna to the receive SDR.
Fig.~\ref{fig:SDRExpeSetup} shows the indoor experimental setup whereby the fabricated ER-FAS serves as the Tx, while a standard horn antenna (SH2000) is utilized as the Rx. For benchmarking the performance of the proposed ER-FAS system, the following precautions are taken to minimize the random small-scale fading due to multipath: (i) both Tx and Rx are placed in direct line-of-sight of each other, and away from all kinds of surrounding objects, (ii) a relatively small transmit power budget is allocated to the transmit node, (iii) each of the Tx and Rx is placed on a table covered by absorbers to avoid reflection from the ground. Under this experimental setup, two distinct experiments are conducted to do a performance comparison of the conventional antenna and ER-FAS by measuring the following at the receiver: (i) received power, and (ii) BER. 

\subsubsection{Received Signal Power}
The first experiment aimed to measure the received signal level (RSL) at the receive SDR under two scenarios: i) when the transmit SDR utilized a conventional antenna, and ii) when it utilized the ER-FAS. In line with the simulation and measurement setup in previous sections, experiments are performed with $-90^\circ$ and $60^\circ$ beamforming directions to thoroughly examine the impact of ER-FAS on signal reception. The Tx sent 4-QAM data at a fixed transmit power of 5 mW using both conventional antenna and ER-FAS, while the receiver is placed at 4 different near-field distances (0.3 m, 0.6 m, 0.9 m, and 1.2 m). This allows a steady decrease in received power at the received side to be observed for both scenarios, i.e., conventional antenna and ER-FAS employed at the transmit side. This is in turn due to a monotonic increase in pathloss as the Tx-Rx separation $d_\mathrm{TR}$ increases (Fig.~\ref{fig:com90}-(a) and Fig.~\ref{fig:com60}-(a)). This is a physically consistent result as it follows the expected path loss characteristics of wireless signal propagation. Secondly, for both $-90^\circ$ and $60^\circ$ beamforming directions, the ER-FAS significantly enhanced the received power compared to the conventional antenna, basically due to the directional gain offered by the ER-FAS. 

 {The ER-FAS prototype exhibits an average received power enhancement of approximately 4-6 dB at large beamforming angles. As the noise bandwidth and receiver front end remain unchanged, the RSL improvement directly increases the SNR gain. In our measurement environment, the noise power is approximately \unit[-95]{dBm}. This indicates that the fabricated ER-FAS prototype can enhance the signal-to-noise ratio (SNR) of the received signal from \unit[53]{dB} to \unit[60]{dB} when $d_\mathrm{TR}=\unit[1.2]{m}$ at $\theta_\mathrm{az}=\unit[90]{^\circ}$. Typically, a higher SNR can result in lower error probability and enable a higher feasible modulation order, which is a fundamental relationship that can be found in, e.g.,~\cite[Fig.~4.7]{heath2018foundations}. Table~\ref{tab:MCSmap} summarizes the typical relationship between the modulation order, SNR requirement, and error vector magnitude (EVM) target under various modulation and coding schemes (MCSs). As demonstrated in Table~\ref{tab:MCSmap}, a \unit[7]{dB} SNR enhancement introduced by our ER-FAS prototype can potentially enable the use of higher-order modulations, e.g., from QPSK to 16-QAM or from 16-QAM to 64-QAM. This demonstrates the capability of our design to significantly improve spectral efficiency and support higher data rates under the same channel and power conditions. }

\begin{table}[t]
\centering
\caption{  {SNR and EVM requirements for LTE/NR MCSs. SNR from Fig.~4.7 of~\cite{heath2018foundations}, EVM from Table~6.5.2.2-1 of 3GPP TS 38.104~\cite{3GPP38104}.}}
\setlength{\tabcolsep}{4pt}
\renewcommand{\arraystretch}{1.2}
  {
\begin{tabular}{c c c c}
\hline
\textbf{LTE MCS} & \textbf{Modulation} & \textbf{Minimum SNR (dB)} & \textbf{EVM (\%)} \\
\hline
2  & QPSK     & -6.0  & 17.5 \\
5  & QPSK     & -2.6  & 17.5 \\
8  & QPSK     & 0.0   & 17.5 \\
11 & 16-QAM   & 3.0   & 12.5 \\
14 & 16-QAM   & 5.0   & 12.5 \\
17 & 16-QAM   & 7.4   & 12.5 \\
20 & 16-QAM   & 10.0  & 12.5 \\
23 & 64-QAM   & 12.3  & 8.0 \\
26 & 64-QAM   & 16.4  & 8.0 \\
\hline
\end{tabular}
}
\label{tab:MCSmap}
\end{table}

\subsubsection{Communication Bit Error Rate}
As for the second experiment, BER measurements are conducted for both the conventional antenna and ER-FAS under $-90^\circ$ and $60^\circ$ beamforming setups. Specifically, an image (of size $162\times 311\times 3$) was transferred using 4-QAM from the transmit side. The transmit power $P$ varies from 0.2 to 0.51 mW for 1.2 m distance and 0.45 to 4.05 mW for 5 m distance to systematically adjust the channel conditions and observe the corresponding changes in the BER. 
The results, depicted in Fig.~\ref{fig:com90}-(b)(c) and Fig.~\ref{fig:com60}-(b)(c), illustrate the relationship between BER and the transmit power (in mW).
For both beamforming configurations (i.e., $-90^\circ$ and $60^\circ$), the ER-FAS consistently outperformed the conventional antenna by exhibiting a lower BER across nearly the full range of transmit power values. Further, the BER improvement brought by the proposed ER-FAS system is more pronounced for the medium and lower values of the transmit power, which correspond to the low-to-medium SNR conditions. This highlights the capability of the proposed ER-FAS system in combating poor channel conditions.

\section{Discussion}\label{sec:Discussion}

\subsection{Comparison with Prior Works}
The main novelty and contributions of this work, compared with previously published studies, are summarized in this section.

On the one hand, from the communication modeling perspective, prior theoretical studies on FAS have primarily focused on spatial reconfigurability (i.e., SR-FAS)~\cite{WKK2021First_Paper_on_FAS, ChaoWang2024FAS, NWK2025Oversampling, WKK2023FAS_Part1, WKK2023FAS_Part3, Khammassi2023TWC, WKK2020FAS_limit}. In SR-FAS, the antenna position (i.e., port) is physically reconfigured~\cite{WKK2023FAS_Part1}, which share the same concept of movable antennas that introduce spatial degrees of freedom to enhance wireless systems~\cite{Zhu_lipeng2024Magzine_MA, zhu2024historicalreviewfluidantenna}. In contrast, this work proposes the concept of Electromagnetically Reconfigurable FAS (ER-FAS), which introduces the electromagnetic degree of freedom while maintaining a fixed antenna position or port. Both far-field and near-field channel models are established, and a beamforming algorithm is developed under this new framework.

On the other hand, from the hardware perspective, four representative liquid-based FAS designs~\cite{Song_lingnan2019TAP, WKK2022FASHardware, Yang2020TAP, Ren2025TAP} and one pixel-based FAS design~\cite{Zhang_jichen2025OJAP} are comprehensively compared in Table~\ref{Comparison Table}. A key distinction of the proposed ER-FAS lies in its element-level reconfigurability, meaning that each array element is independently reconfigurable. Thus, the ER-FAS may also be regarded as an element-reconfigurable array (ERA). In contrast, previously published works~\cite{WKK2022FASHardware, Yang2020TAP, Ren2025TAP, Zhang_jichen2025OJAP} achieve radiation pattern control at the whole-antenna level. Furthermore, those designs are typically bulky (larger than half a wavelength), leading to grating lobes and limiting their suitability for large-scale arrays. In contrast, the proposed ER-FAS supports large-angle far-field beamscanning and near-field beamfocusing, enhancing wireless performance. Additional comparisons in terms of reconfiguration mechanism, beam scanning range, wireless channel modeling, and system-level functions are also provided in Table~\ref{Comparison Table}.

\begin{table*}[t]
  \renewcommand{\arraystretch}{1.4}
  \begin{center}
  \caption{Antenna System Comparison with Previously Reported Works}
  \label{Comparison Table}
  \scriptsize
  \begin{tabular}{
    >{\centering\arraybackslash}p{1.6cm} !{\vrule width1pt} 
    >{\centering\arraybackslash}p{2.1cm} !{\vrule width1pt} 
    >{\centering\arraybackslash}p{1.6cm} !{\vrule width1pt} 
    >{\centering\arraybackslash}p{2.6cm} !{\vrule width1pt} 
    >{\centering\arraybackslash}p{2.2cm} !{\vrule width1pt} 
    >{\centering\arraybackslash}p{2.6cm} !{\vrule width1pt} 
    >{\columncolor{black!8}\centering\arraybackslash}p{2.9cm}
  }
    \Xhline{1pt}
    Ref. & ~\cite{Song_lingnan2019TAP} & ~\cite{WKK2022FASHardware} & ~\cite{Yang2020TAP} & ~\cite{Ren2025TAP} & ~\cite{Zhang_jichen2025OJAP} & This work \\
    \Xhline{1pt}
    Type \& Technique & Liquid-based; Patch antenna with switchable slots & Liquid-based; Surface-wave fluid antenna & Liquid-based; Fabry–Pérot antenna with liquid-metal partially reflective surface & Liquid-based; Feeding antenna with liquid dielectric & Pixel-based; E-slot patch with upper pixels & \textcolor{black}{Liquid-based; Directors and reflectors with planar monopole} \\
    \hline
    Antenna size ($\lambda \times \lambda$) & 0.8 × 0.8 & 3.3 × 1.0 & 1.6 × 1.6 & 1.48 × 1.48 & 0.67 × 0.67 & \textcolor{black}{0.5 × 0.5} \\
    \hline
    Array configuration feasibility & No & No & No & No & No & \textcolor{black}{Yes, 1 × 12 array is constructed (can be extended to 2D array)} \\
    \hline
    Reconfigurable mechanism & Liquid metal with printed microfluidics and syringes & Electrowetting &  Liquid metal with syringes and rubber tube & Liquid dielectric with printed pipeline & PIN diode & \textcolor{black}{Liquid metal controlled by pumping machine} \\
    \hline
    Beam scanning & Fixed pattern & $30^\circ \sim 60^\circ$ & $-20^\circ \sim 20^\circ$ & 3 discrete states & 12 discrete states & \textcolor{black}{$-90^\circ \sim 90^\circ$} \\
    \hline
    Wireless channel modeling & No & No & No & No & Yes, modeling the correlation between FAS ports and antenna patterns & \textcolor{black}{Yes, both near and far fields, and joint beamforming algorithm} \\
    \hline
    Functions & Wide frequency tuning & Agile radiation patterns & Antenna radiation pattern and 3 dB beamwidth reconfigurability & For satellite antennas at three different orbit altitudes & 12 radiation patterns control with high swithcing speed & \textcolor{black}{Far-field large-angle beamscanning; near-field beamfocusing} \\
    \hline
    Main contributions & Wideband frequency reconfigurability with consistent radiation patterns & Experimentally validate the travelling speed of the FAS utilizing CEW & Separately control antenna beam direction and beamwidth & Three patterns of isoflux, flat-top, and pencil-beam patterns with circular polarization & Validate the pattern diversity and the port correlations of the pixel-based FAS in system level & \textcolor{black}{ER-FAS concept, channel modeling, and beamforming algorithm; ER-FAS hardware design, prototype, and experiment in array configuration} \\
    \Xhline{1pt}
  \end{tabular}
  \end{center}
\end{table*}

\subsection{Future Research Directions}

This work systematically demonstrates an ER-FAS as a proof of concept, with several promising directions for further research. The pattern-reconfigurable ER-FAS can support advanced communication models and algorithms, such as tri-hybrid multi-user precoding. From the hardware side, future extensions may incorporate reconfigurable phase shifters and integrated feeding networks to enable adaptive or hybrid beamforming in dynamic environments. Additionally, power dividers and phase shifters may be implemented using fluidic mechanisms for unified system control. The current validation is performed under a static setup to ensure clean and repeatable evaluation of radiation pattern control. Future developments may extend the system to dynamic scenarios with mobile receivers or time-varying channels to assess real-time adaptability and robustness.
 
Additionally, the proposed ER-FAS concept can be extended to achieve other forms of electromagnetic reconfigurability. One direction is frequency reconfigurability, which can be realized by tuning the effective length of the fluidic radiating structure—by injecting liquid metal into variable-length channels or altering the geometry of the fluid reservoir. Another promising type is polarization reconfigurability, which can be enabled by introducing asymmetric fluid-metal channel configurations for the excitation monopole or by employing orthogonally arranged reconfigurable arms in the director and reflector layers. These approaches may allow switching between different linear polarizations or between linear and circular polarizations. Although these directions are beyond the scope of the present study, they represent promising future extensions of the ER-FAS platform.

\section{Conclusion}\label{sec:Conclusion}

In this work, a novel design of the electromagnetically reconfigurable fluid antenna system (ER-FAS) is proposed for wireless communication applications, where each ER-FAS array element can be electromagnetically reconfigured. We present the entire process, including electromagnetic design, wireless channel modeling, beamforming algorithm development, full-wave simulation, hardware fabrication, and experimental validation. The results demonstrate that the designed ER-FAS effectively addresses the large-angle beam-scanning issue in the far field and achieves better beam-focusing performance in the near field, thus enhancing the spectral efficiency of the overall wireless communication system.

\bibliography{references}
\bibliographystyle{IEEEtran}

\end{document}